\documentclass[twocolumn]{aastex63}
\usepackage{amsmath}

\newcommand{\ha}{\hbox{H$\alpha$}}

\newcommand{\hii}{\hbox{H\,{\sc ii}}}

\newcommand{\gsim}{\lower.5ex\hbox{$\; \buildrel > \over \sim \;$}}
\newcommand{\lsim}{\lower.5ex\hbox{$\; \buildrel < \over \sim \;$}}
\newcommand{\oi}{\hbox{[O\,{\sc i}]}}
\newcommand{\oii}{\hbox{[O\,{\sc ii}]}}
\newcommand{\oiii}{\hbox{[O\,{\sc iii}]}}
\newcommand{\nii}{\hbox{[N\,{\sc ii}]}}
\newcommand{\sii}{\hbox{[S\,{\sc ii}]}}

\received{M D, 2020}
\revised{M D, 2020}
\submitjournal{ApJ}

\shorttitle{Te based local scaling relationship}
\shortauthors{Yao et al.}

\begin{document}

\title{Sub-galactic scaling relations with T$_{\rm e}$-based metallicity of low metallicity regions in galaxies: metal-poor gas inflow may have important effects?}

\correspondingauthor{Xu Kong}
\email{yaoyao97@mail.ustc.edu.cn}\email{xkong@ustc.edu.cn}

\author[0000-0002-6873-8779]{Yao Yao}
\affiliation{CAS Key Laboratory for Research in Galaxies and Cosmology, Department of Astronomy, University of Science and Technology of China, Hefei 230026, People’s Republic of China}
\affiliation{School of Astronomy and Space Science, University of Science and Technology of China, Hefei 230026, People’s Republic of China}
\author{Haiyang Liu}
\affiliation{CAS Key Laboratory for Research in Galaxies and Cosmology, Department of Astronomy, University of Science and Technology of China, Hefei 230026, People’s Republic of China}
\affiliation{School of Astronomy and Space Science, University of Science and Technology of China, Hefei 230026, People’s Republic of China}
\author{Xu Kong}
\affiliation{CAS Key Laboratory for Research in Galaxies and Cosmology, Department of Astronomy, University of Science and Technology of China, Hefei 230026, People’s Republic of China}
\affiliation{School of Astronomy and Space Science, University of Science and Technology of China, Hefei 230026, People’s Republic of China}
\author[0000-0002-5973-694X]{Yulong Gao}
\affiliation{Department of Astronomy, Nanjing University, Nanjing 210093, China}
\affiliation{Key Laboratory of Modern Astronomy and Astrophysics (Nanjing University), Ministry of Education, Nanjing 210093, China}
\author{Guangwen Chen}
\affiliation{CAS Key Laboratory for Research in Galaxies and Cosmology, Department of Astronomy, University of Science and Technology of China, Hefei 230026, People’s Republic of China}
\affiliation{School of Astronomy and Space Science, University of Science and Technology of China, Hefei 230026, People’s Republic of China}
\author{Xinkai Chen}
\affiliation{CAS Key Laboratory for Research in Galaxies and Cosmology, Department of Astronomy, University of Science and Technology of China, Hefei 230026, People’s Republic of China}
\affiliation{School of Astronomy and Space Science, University of Science and Technology of China, Hefei 230026, People’s Republic of China}
\author{Zhixiong Liang}
\affiliation{CAS Key Laboratory for Research in Galaxies and Cosmology, Department of Astronomy, University of Science and Technology of China, Hefei 230026, People’s Republic of China}
\affiliation{School of Astronomy and Space Science, University of Science and Technology of China, Hefei 230026, People’s Republic of China}
\author{Zesen Lin}
\affiliation{CAS Key Laboratory for Research in Galaxies and Cosmology, Department of Astronomy, University of Science and Technology of China, Hefei 230026, People’s Republic of China}
\affiliation{School of Astronomy and Space Science, University of Science and Technology of China, Hefei 230026, People’s Republic of China}
\author{Yimeng Tang}
\affiliation{CAS Key Laboratory for Research in Galaxies and Cosmology, Department of Astronomy, University of Science and Technology of China, Hefei 230026, People’s Republic of China}
\affiliation{School of Astronomy and Space Science, University of Science and Technology of China, Hefei 230026, People’s Republic of China}
\author{Hongxin Zhang}
\affiliation{CAS Key Laboratory for Research in Galaxies and Cosmology, Department of Astronomy, University of Science and Technology of China, Hefei 230026, People’s Republic of China}
\affiliation{School of Astronomy and Space Science, University of Science and Technology of China, Hefei 230026, People’s Republic of China}


\begin{abstract}

The scaling relationship is a fundamental probe of the evolution of galaxies. Using the integral field spectroscopic data from the Mapping Nearby Galaxies at Apache Point Observatory survey, we select 1698 spaxels with significant detection of the auroral emission line \oiii$\lambda$4363 from 52 galaxies to investigate the scaling relationships at the low-metallicity end. We find that our sample's star formation rate is higher and its metallicity is lower in the scaling relationship than the star-forming sequence  after removing the contribution of the Fundamental Metallicity Relation.We also find that the stellar ages of our sample are younger ($<$ 1 Gyr)  and the stellar metallicities are also lower. Morphological parameters from Deep Learning catalog indicate that our galaxies are more likely to be merger.  These results suggest that their low metallicity regions may be related to interaction, the inflow of metal-poor gas may dilute the interstellar medium and form new metal-poor stars in these galaxies during interaction.

\end{abstract}



\section{Introduction} \label{sec:introduction}
The scaling relationship is important for understanding the formation and evolution of galaxies. One of the well-established relationships is the correlation between the star formation rate (SFR) and the stellar mass (M$_*$), which is the so-called star-forming main sequence (SFMS). Another important relationship is the stellar mass–metallicity relation (MZR), which indicates that galaxy metallicities increase with increasing M$_*$, and reflects the balance between gravitational potential and galactic feedback. It was established by \cite{1979A&A....80..155L} and has been extended by a series of studies for decades \citep[e.g., ][]{2004ApJ...613..898T,2010MNRAS.408.2115M}.

The SFMS and MZR are both established primarily by global galactic parameters. It is also important to understand whether galactic local parameters (e.g., stellar mass surface density ($\Sigma_*$), SFR surface density ($\Sigma_{\rm SFR}$), and local metallicity) are more fundamental to probe the global SFMS and MZR. Recently, with the emergence of integral field spectroscopy (IFS) surveys, spatially resolved scaling relationships have also been developed. \cite{2012ApJ...756L..31R} demonstrated the existence of a local relation between $\Sigma_*$, metallicity, and $\Sigma_{\rm SFR}$ using 38 nearby galaxies from the PPAK IFS Nearby Galaxies Survey \citep{2010MNRAS.405..735R} and the Calar Alto Legacy Integral Field spectroscopy Area \citep[CALIFA;][]{2012A&A...538A...8S} survey. \cite{2016ApJ...821L..26C} found a spatially resolved SFMS on $\sim$1 kpc scale with a slope of $\sim$0.7 in the local universe from CALIFA IFS data. More further researches on the spatially resolved scaling relationship also followed on (e.g.  \citealt{2016MNRAS.463.2513B}; \citealt{2018ApJ...868...89G}, hereafter G18; \citealt{2018ApJ...857...17L}).

The metal-poor galaxies/regions play an essential role in galaxy evolution, especially for the galaxies at the early stage of evolution, or for those that evolve slowly. Understanding the various properties of galaxies at low metallicity in the MZR can help us uncover some important processes in galaxy evolution. Low metallicities of galaxies/regions suggest that they might have significant metal-poor gas inflows or metal-enriched gas outflows. \citep{2000A&ARv..10....1K}. However, the slope and scatter of the low metallicity end of the local MZR is still not clear currently. The reason is that current methods of investigating the local MZR are mainly based on the strong emission lines method, which may be not accurate enough for the calibration of low metallicity \citep{2011MNRAS.411.2076L}. The electron temperature ($T_{\rm e}$) method \citep{1984ASSL..112.....A,2006A&A...448..955I,2010ApJ...720.1738P}, which is also called direct method, base on the anticorrelation between metallicity and $T_{\rm e}$, can obtain the metallicity directly and avoid the systematic error of low metallicity  with high ionization parameter. It needs auroral line ratios such as \oiii$\lambda$4363/\oiii$\lambda\lambda$4959,5007 to calculate $T_{\rm e}$. Auroral lines are two order of magnitude fainter than strong lines, which makes the direct method very challenging to apply. 
But with the sample size of the current IFS surveys rapidly expanding recent years, we finally have the dataset to apply  direct method  by searching for resolved \oiii$\lambda$4363 emission lines for a large and representative sample of galaxies.

In this work, we select \hii{} regions with significant \oiii$\lambda$4363 emission line dectection from SDSS-IV Mapping Nearby Galaxies at Apache (MaNGA) survey \citep{2015ApJ...798....7B} and present their local SFMS and MZR. The paper is organized as follows. Section \ref{sec:data} describes the sample selection and the calculation of physical parameters. Section \ref{sec:result} presents our main results. Section \ref{sec:discussion} discusses influences of metallicity calibrations and the potential role of merges in our sample. Finally, Section \ref{sec:conclusion} comes up with our conclusions. We adopt a flat cosmology with $\Omega_\Lambda$ = 0.7, $\Omega_M$ = 0.3 and $H_0$ = 70 km s$^{-1}$ Mpc$^{-1}$ to determine distance-dependent measurements.

\section{Data} \label{sec:data}

\subsection{MaNGA DR15 Overview} \label{subsec:manga}
The MaNGA survey, one of the three core programs in the SDSS-IV \citep{2017AJ....154...28B}, is an IFS survey targeted at $\sim10,000$ nearby galaxies that are selected from the NASA-Sloan-Atlas catalog\footnote{\url{http://www.nsatlas.org}} \citep[NSA;][]{2011AJ....142...31B, 2015ApJ...798....7B}. The redshifts of these target galaxies span a range of $0.01<z<0.15$. The spectral coverage is 3600–10300 \AA{} with a resolution of R $\sim$ 2000 \citep{2015AJ....149...77D}. The full width at half maximum (FWHM) of the reconstructed point-spread function is about 2\farcs5 \citep{2016AJ....152...83L}. Recently, the SDSS DR15 publicly released IFS observations and ancillary data products of 4621 galaxies \citep{2019ApJS..240...23A}. Comparing with DR14, there are many improvements in DR15, with Data Reduction Pipeline \citep[DRP;][]{2016AJ....152...83L}, such as the flux calibration and spetral line-spread function estimation. DR15 also provides an entirely new Data Analysis Pipeline products \citep[DAP;][]{2019AJ....158..231W}\footnote{\url{https://www.sdss.org/dr15/manga/manga-analysis-pipeline/}}, which provides the measurements of Balmer series lines and strong forbidden lines, including \oii$\lambda$3727, \oiii$\lambda\lambda$4959,5007, and \nii$\lambda$6583 \citep{2019AJ....158..160B}, facilitating our data analysis. In this work, we treat these 4621 galaxies from DRP and DAP as the parent sample.

\subsection{Rough Sample Selection} \label{subsec:roughselection}
In order to identify regions with strong \oiii$\lambda$4363 from all MaNGA data cubes more efficiently , we first select spaxels according to the following criteria:

\begin{enumerate}
\item $b/a\geqslant 0.3$,
\item no ``bad'' flag in ``DRP3QUAL'',
\item S/N(H$\alpha$) $\geqslant$ 3, S/N(H$\beta$) $\geqslant$ 3, S/N(\oiii$\lambda$3727) $\geqslant$ 3, S/N(\oiii$\lambda\lambda$4959,5007) $\geqslant$ 3, S/N(\nii$\lambda$6583) $\geqslant$ 3, and S/N(\oiii$\lambda$4363) $\geqslant$ 5,
\item not active galactic nuclei (AGN).
\end{enumerate}

The values of $b$/$a$ are from the NSA catalog, representing the projected ratio of the semi-minor axis to the semi-major axis of galaxies. This limitation is to avoid the severe dust attenuation in galactic disks. The fluxes and noises of all emission lines except \oiii$\lambda$4363 are derived based on the DAP hybrid bin, while the fluxes and noises of \oiii$\lambda$4363 are fitted to a single Gaussian profile by ourselves following \cite{2014ApJ...780..122L}. All data of emission lines have been processed by Galactic extinction correction, moving to rest frame, and intrinsic dust extinction correction in turn. We use the color excess $E(B-V)$ map of the Milky Way \citep{1998ApJ...500..525S} and the \cite{1989ApJ...345..245C} extinction law to correct Galactic extinction and use the Balmer decrements under the Case B recombination and apply the \cite{2000ApJ...533..682C} attenuation law to correct intrinsic dust extinction. For spaxels with H$\alpha$/H$\beta<$2.86 \citep{1987MNRAS.224..801H}, we assume the extinction as zero. The \nii-based BPT diagnostic diagram \citep{1981PASP...93....5B} is adopted to avoid possible AGN contamination, and only spaxels below the \cite{2003MNRAS.346.1055K} demarcation curve are selected. Thus, we obtain a rough selected sample with 3565 spaxels.

\subsection{Spectral Fitting and Further Sample Selection} \label{subsec:specfit}

In Section \ref{subsec:roughselection}, to search for spaxels with obvious \oiii$\lambda$4363 emission line quickly, we fit the spectra with single Gaussian profile without subtracting stellar continuum. So the selection is rough and is needed to be refined. In this Section, we refit the stellar continuum and emission lines for our rough selected sample.

Following the procedure of \cite{2017RAA....17...41G}, we use the \texttt{STARLIGHT} software\footnote{Available at \url{http://www.starlight.ufsc.br/downloads/}} \citep{2005MNRAS.358..363C} to obtain the best-fit stellar continua, which are then subtracted from the observed spectra. The fitting is based on the built-in stellar library \citep{2003MNRAS.344.1000B} of \texttt{STARLIGHT} and a \cite{2003PASP..115..763C} initial mass function (IMF). The stellar library includes 45 single stellar populations, consisting of 15 stellar ages ($t_*$, ranging from 1 Myr to 13 Gyr) and 3 different stellar metallicities ($Z_*$ = 0.004, 0.02, and 0.05). Several stellar properties, such as M$_*$, $Z_*$, and $t_*$, can be derived from the output of \texttt{STARLIGHT}. In order to ensure the consistency of flux measurements, we use the \texttt{MPFIT} \citep{2009ASPC..411..251M} to refit emissions lines that we are interested in (e.g., \oii$\lambda$3727, \oiii$\lambda$4363, H$\beta$, \oiii$\lambda\lambda$4959,5007, H$\alpha$, \nii$\lambda\lambda$6548,6583, \sii$\lambda\lambda$6717,6731). After obtaining the fluxes of emission lines, we apply Balmer decrement \citep{1987MNRAS.224..801H} and \cite{2000ApJ...533..682C} 
attenuation curve to them again. Finally, we get the final emission fluxes of this work, rather than the DAP data and the rough flux measurements in Section \ref{subsec:roughselection}. {\bf In order to verify the accuracy of our fitting, we compared our fitting results with the DAP results. Details of the comparison are in Appendix \ref{sec:comparison_dap}.}

Because \oiii$\lambda$4363 emission might be contaminated by cosmic rays or other noise, further selection is needed to remove fake detection. First of all, we remove galaxies whose spaxel numbers in the rough selected sample are less than 5. Then, we calculate the S/N ratio at $\lambda$ = 4020 \AA\ and remove the spaxels with S/N$<$5 to ensure that the uncertainty of stellar mass are less than 0.11 dex \citep{2005MNRAS.358..363C}. Spaxels whose fluxes of \oiii$\lambda$4363 are too excessive to calculate an vaild metallicity are also removed, which may be caused by noise. Next, we manually check the continuum fitting status and emission line fitting status of each spectrum. We remove those spectra whose \oiii$\lambda$4363 lines do not have well-defined profiles or velocities have significant deviation from other emission lines. After manual inspection, our final sample consists of 1698 spaxels from 52 different galaxies.



In addition, we also select all star-forming regions in these 52 galaxies. The requirements of S/N are the same as those of the \oiii$\lambda$4363 regions without the \oiii$\lambda$4363 emission line to calculate their metallicities applying the strong line method.


\subsection{SFR Surface Density and Metallicity Calculation} \label{subsec:furtherselection}
\subsubsection{SFR Surface Density}
We use the H$\alpha$ luminosity to determine the dust-corrected SFR for each spaxel, assuming a \cite{2003PASP..115..763C} IMF. The SFR calibration from \cite{1998ARA&A..36..189K} is:

\begin{equation}\label{eq:sfr_calc}
{\rm SFR(M_{\odot}~yr^{-1})}=4.4 \times 10^{-42} L_{\mathrm{H}\alpha}\mathrm{(erg~s^{-1})}.
\end{equation}
The size of each spaxel of MaNGA is $0\farcs5 \times 0\farcs5$, so the projection corrected area is given by

\begin{equation}
A=\left[D(z) \times 0.5 \times \frac{\pi}{3600 \times 180}\right]^2 \times \frac{1}{b/a},
\end{equation}
where $D(z)$ is the angular diameter distance of the galaxy. As a result, the $\Sigma_{\rm SFR}$ and $\Sigma_*$ of each spaxel are

\begin{equation}
{\rm \Sigma_{SFR} = SFR/}A,
\end{equation}

\begin{equation}
\Sigma_*=M_*/A.
\end{equation}

\subsubsection{$T_{\rm e}$ and Oxygen abundence}
Here, we adopt the formulation from \cite{2010ApJ...720.1738P} to calculate the $T_{\rm e}$, O$^{2+}$ abundance and O$^+$ abundance of each spaxel. For our metallicity estimation, we adopt a standard two-zone temperature model with $t_2$ = 0.672 $t_3$ + 0.314 \citep{2009MNRAS.398..485P}. Because the most important ions of oxygen in \hii{} regions are O$^+$ and O$^{2+}$, we take the sum of the abundance of these two ions as the abundance of oxygen.

To all star-forming spaxels without the requirement of S/N(\oiii$\lambda$4363)$\geqslant$3 which are described at the end of Section \ref{subsec:specfit}, we apply the R calibration relation from \cite{2016MNRAS.457.3678P}(hereafter PG16), which is calibrated from the T$_{\rm e}$ metallicities of 313 \hii{} regions, ranging from 12+log(O/H)$\sim$7.0 to $\sim$8.7. with a mean difference of 0.049 dex.

\subsection{Uncertainty Estimate} \label{subsec:uncertainty}
Our uncertainty consists of two parts, part one is converted from the error spectrum provided by MaNGA DATACUBE ($\sigma_{\rm errspec}$) , and part two comes from 2 independent observations of the same galaxy (8256-9102 and 8274-9102) in our sample ($\sigma_{\rm indepobs}$). Our final uncertainty $\sigma$ is determined by rooting the sum of the squares of the above two parts: 

\begin{equation}
\sigma=\sqrt{\sigma_{\rm errspec}^2+\sigma_{\rm indepobs}^2}.
\end{equation}

For uncertainties of $\Sigma_{\rm SFR}$ and 12+log(O/H) from the error spectrum ($\sigma_{\rm errspec}$), we adopt a Monte-Carlo method and repeat the calculation for 1000 times. For every spectrum, we produce a series of fluxes for each emission line with a Gaussian distribution, assuming its average as the measured line flux and standard deviation as the measured error. The median of metallicity uncertainty is {\bf 0.07} dex, and the median of $\Sigma_{\rm SFR}$ uncertainty is {\bf 0.008} dex. For uncertainties of $\Sigma_{*}$ from the error spectrum, we directly adopt 0.11 dex \citep{2005MNRAS.358..363C}, because the S/N ratios of our continuua are all grater than 5.

For uncertainties from 2 independent observations, we can get a distribution by taking the difference of the quantities at the same coordinates. We use the standard deviation of this distribution as the uncertainty. The standard deviations ($\sigma_{\rm indepobs}$) of $\Sigma_{\rm SFR}$, $\Sigma_{*}$ and metallicity are 0.10 dex, 0.17 dex and 0.10 dex, respectively. The final uncertainty ($\sigma$) of $\Sigma_{\rm SFR}$, $\Sigma_{*}$ and metallicity are 0.10 dex, 0.20 dex and {\bf 0.12} dex, respectively.

\section{Result} \label{sec:result}
Comparing to previous studies \citep[e.g.,][]{2016MNRAS.463.2513B, 2018ApJ...868...89G}, this work uses the newest DR15 data as the initial sample and focuses on the \oiii$\lambda$4363 selected star-forming regions. Because of the weakness of \oiii$\lambda$4363 emission lines, our sample is much smaller than normal star forming sample, and its overall properties are also much different form those of normal star forming sample. The SDSS images and metallicity profiles of 4 galaxies with the most spaxels in our 52 \oiii$\lambda$4363 selected sample are shown in Figure \ref{fig:metal_profile_0_5}. Other galaxies are shown in Appendix \ref{sec:other_galaxies}. We perform a linear fit to the profile of the metallicity of each galaxy along the radius.

\begin{figure*}[ht!]
	\centering
	\includegraphics[width=1\textwidth]{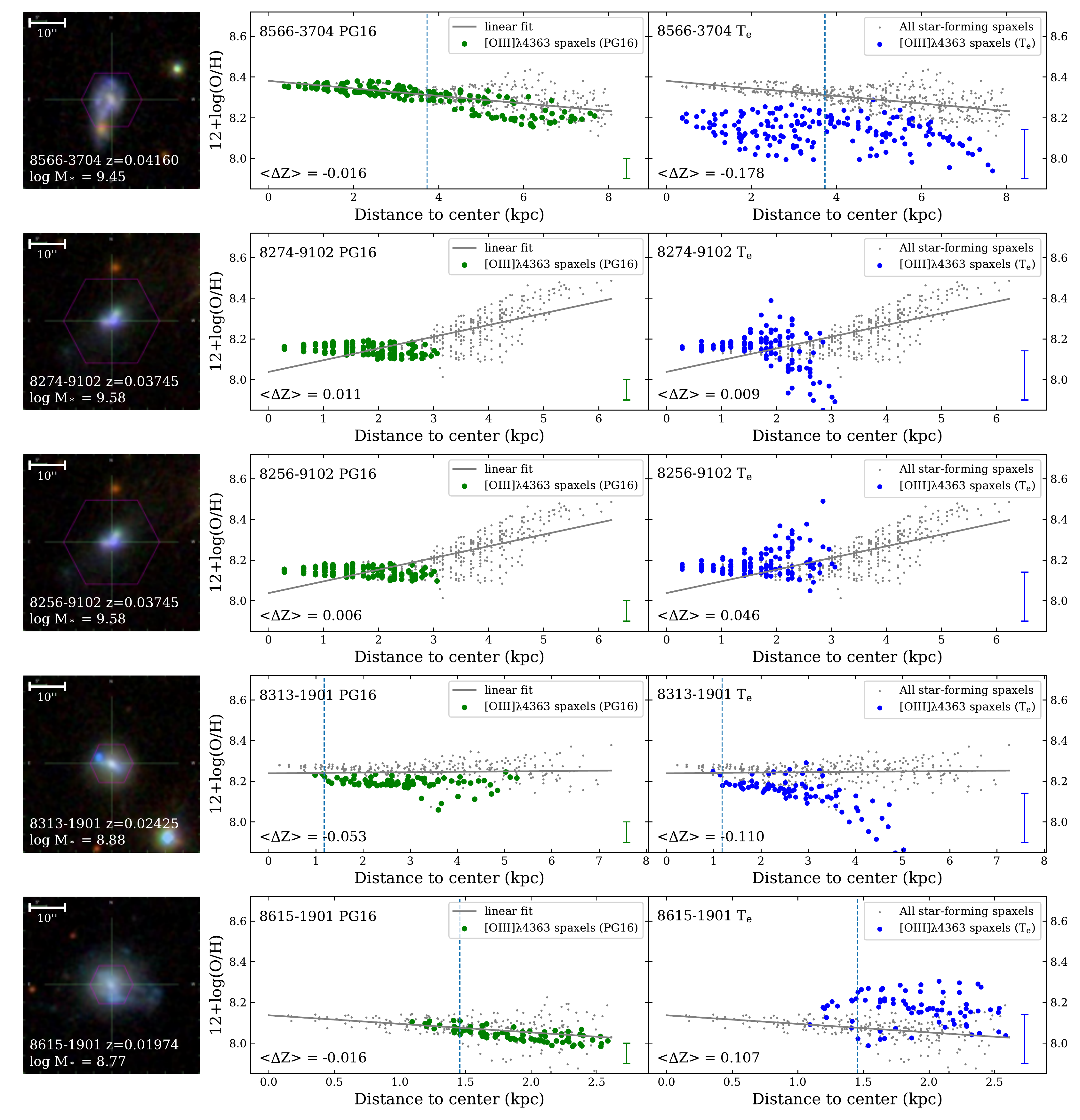}
	\caption{Left panel: the SDSS images of 5 galaxies with the most spaxels in our 52 \oiii$\lambda$4363 selected sample. The field of view of each image is 50\arcsec. Right panel: the radial profiles of the metallicity of the corresponding galaxy. Gray dots are all star-forming spaxels without the requirement of S/N(\oiii$\lambda$4363)$\geqslant$3, which are described at the end of Section \ref{subsec:specfit}. Gray lines are the linear fit of gray dots. Green dots and blue dots represent the PG16 and T$_{\rm e}$ calibrated metallicity of our \oiii$\lambda$4363 selected spaxels, respectively. The vertical dashed line represents the effective radius of the galaxy. Both 2 independent observations of the same galaxy (8274-9102 and 8256-9102) are shown in the second and third row of the figure, respectively. \label{fig:metal_profile_0_5}}
\end{figure*}

\subsection{Global SFMS and Morphology} \label{subsec:global_result}

\begin{deluxetable*}{cccccccccc}
	\tablenum{1}
	\tablecaption{Basic information of 30 galaxies with the most spaxels in our sample\label{tab:galaxies_info}}
	\tablewidth{0pt}
	\tablehead{
		\colhead{Galaxy ID} & \colhead{Count} & \colhead{RA} & \colhead{DEC} & \colhead{Redshift} & \colhead{M$_*$} & \colhead{M$_*$} & \colhead{SFR} & \colhead{T-Type} & \colhead{P(merge)}\\
		\colhead{[plate]-[ifu]} & \colhead{} & \colhead{(deg)} & \colhead{(deg)} & \colhead{} & \colhead{(log(M${_*}$/M$_\odot$))} & \colhead{(log(M${_*}$/M$_\odot$))}& \colhead{(log(M$_\odot$/yr))} & \colhead{} & \colhead{}
	}
	\decimalcolnumbers
	\startdata
	8566-3704 & 173 & 115.22481 & 40.06964 & 0.04160 & 9.45 & $-$ & $-$ & 4.30 & 0.99 \\
	8274-9102* & 124 & 165.10414 & 43.01969 & 0.03745 & 9.58 & 9.02 & 0.52 & 7.00 & 0.34 \\
	8313-1901 & 84 & 240.28713 & 41.88075 & 0.02425 & 8.88 & 9.28 & -0.04 & 1.90 & 0.18 \\
	8615-1901 & 79 & 321.07220 & 1.02836 & 0.01974 & 8.77 & 8.95 & -0.06 & 4.55 & 0.38 \\
	9000-6102 & 68 & 170.88396 & 53.46820 & 0.02719 & 9.45 & 9.41 & -0.05 & 5.27 & 0.73 \\
	7495-6102 & 62 & 204.51286 & 26.33821 & 0.02615 & 8.74 & 8.96 & 0.12 & 5.80 & 0.06 \\
	8942-3703 & 60 & 124.38507 & 28.35783 & 0.01986 & 8.67 & 8.85 & -0.21 & 4.43 & 0.04 \\
	8551-1902 & 54 & 234.59171 & 45.80194 & 0.02135 & 8.43 & 8.90 & -0.13 & 1.74 & 0.04 \\
	8465-6102 & 48 & 197.54970 & 48.62339 & 0.02828 & 9.29 & 8.93 & 0.40 & 3.08 & 0.96 \\
	8548-9102 & 48 & 244.91463 & 47.87321 & 0.02094 & 8.70 & 8.87 & -0.22 & 5.99 & 0.79 \\
	8455-9101 & 47 & 157.18363 & 39.77859 & 0.03004 & 9.66 & 9.53 & 0.13 & 5.46 & 0.79 \\
	8322-9101 & 44 & 199.60756 & 31.46795 & 0.01869 & 9.54 & $-$ & $-$ & 5.31 & 0.89 \\
	8548-3702 & 43 & 243.32681 & 48.39181 & 0.01990 & 8.74 & 8.87 & -0.44 & 4.47 & 0.45 \\
	8257-3704 & 43 & 165.55361 & 45.30387 & 0.02022 & 8.69 & 9.00 & -0.65 & 10.00 & 0.04 \\
	8719-12702 & 41 & 120.19928 & 46.69053 & 0.01938 & 9.49 & 9.65 & -0.04 & 5.57 & 0.43 \\
	8727-3702 & 41 & 54.55405 & -5.54040 & 0.02215 & 8.48 & 8.76 & -0.16 & -1.16 & 0.04 \\
	8239-12704 & 40 & 117.45125 & 48.12956 & 0.01808 & 9.42 & 9.29 & -0.39 & 6.09 & 0.31 \\
	8241-6101 & 40 & 127.04853 & 17.37466 & 0.02087 & 8.63 & 8.62 & -0.38 & 5.98 & 1.00 \\
	8996-12703 & 37 & 171.97836 & 53.00109 & 0.02134 & 8.84 & 8.13 & -0.29 & 6.59 & 0.78 \\
	8715-12704 & 37 & 121.17245 & 50.71853 & 0.02262 & 8.77 & 8.96 & -0.12 & 6.16 & 0.11 \\
	8987-6103 & 33 & 137.26802 & 28.25683 & 0.02166 & 8.67 & 8.64 & -1.93 & 5.12 & 0.30 \\
	8936-6104 & 32 & 117.92943 & 30.44875 & 0.01424 & 8.65 & 6.99 & -1.90 & 4.64 & 0.09 \\
	8987-9101 & 28 & 137.44943 & 27.86231 & 0.02044 & 8.94 & 8.94 & -0.69 & 4.90 & 0.02 \\
	8552-6101 & 28 & 227.01700 & 42.81902 & 0.01801 & 8.66 & 8.64 & -0.88 & 5.74 & 0.82 \\
	9509-3702 & 27 & 122.43975 & 25.88031 & 0.02511 & 9.27 & 9.70 & 0.45 & 4.65 & 1.00 \\
	8147-9102 & 27 & 118.85282 & 26.98635 & 0.01524 & 8.85 & 8.83 & -0.98 & 5.28 & 0.38 \\
	8553-12703 & 25 & 235.95318 & 57.23323 & 0.01355 & 8.61 & 6.20 & -2.10 & 5.65 & 0.34 \\
	8325-12702 & 24 & 209.89514 & 47.14768 & 0.04204 & 9.28 & 9.50 & 0.43 & 6.35 & 0.17 \\
	8458-3702 & 23 & 147.56250 & 45.95731 & 0.02487 & 9.44 & 7.34 & -0.29 & 3.94 & 0.68 \\
	8083-12705 & 22 & 50.17898 & -1.10865 & 0.02091 & 10.09 & 10.02 & -0.03 & 5.44 & 0.34 \\
	\enddata
	\tablecomments{From left to right, the columns correspond to (1) galaxy ID, defined as the [plate]-[ifudesign] of the MaNGA observations; (2) the count of \oiii$\lambda$4363 spaxels; (3) and (4) the ra and dec of IFU; (5) the redshift taken from NSA catalog; (6) the total M$_*$ taken from NSA sersic mass; (7) the total M$_*$ taken from MPA-JHU catalog, some of galaxies (3 in 52) can not match the catalog, which have a empty value; (8) SFR taken from MPA-JHU catalog, some of galaxies can not match the catalog, which have a empty value; (9) the T-Type value of morphology derived from DL catalog (T-Type < 0 for ETGs, T-Type > 0 for LTGs, T-Type $\sim$ 0 for S0), whose methodology for training and testing the DL algorithm is described in \cite{2018MNRAS.476.3661D}; (10) the probability of merger signature (or projected pair) derived from the DL catalog same as (9).}
	\tablecomments{*: MaNGA has two independent observations (8256-9102 and 8274-9102) of this galaxy, here we only select the observation (8274-9102) with less uncertainty of metallicity.
	}
\end{deluxetable*}

\begin{figure}[ht!]
	\centering
	\includegraphics[width=1\columnwidth]{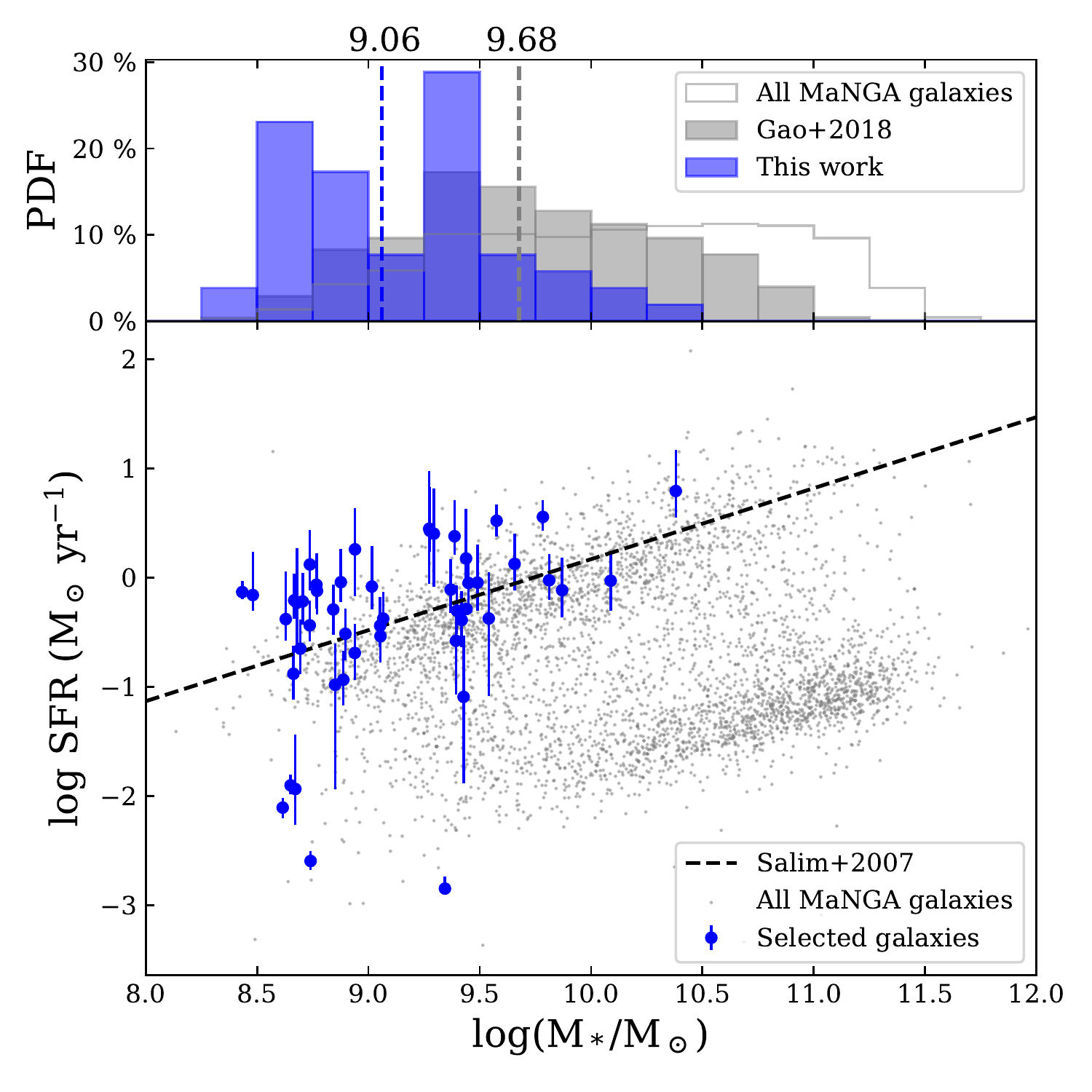}
	\caption{Top panel: the histograms of normalized M$_*$ distribution. The blue filled histograms represent our 52 \oiii$\lambda$4363 emission galaxies, the gray filled histograms represent star forming sample in G18, and the gray hollow histograms all MaNGA galaxies in DRP catalog. The vertical dashed lines of different colors respectively indicate the median of the corresponding mass distribution. Bottom panel: M$_*-$ SFR relation (SFMS) of MaNGA galaxies. Blue dots represent our 50 sample galaxies (4 of 52 galaxies can not be matched in MPA-JHU catalog), whose error bars represent 1$\sigma$ error (16\% $\sim$ 84\%). Gray dots are all of the galaxies in MaNGA DR15 data. The main-sequence relation for SFGs in the local universe \citep{2007ApJS..173..267S} is shown as the black dashed line. \label{fig:global_m_sfr_relation}}
\end{figure}

We first investigate the global properties of these galaxies. Figure \ref{fig:global_m_sfr_relation} shows the mass distribution and the M$_*-$SFR relation (global SFMS) comparing with all MaNGA galaxies and SFGs in G18 which contains 1122 galaxies and 8 $\times$ 10$^5$ spaxels. SFRs are retrieved from the MPA-JHU catalog\footnote{\url{https://www.sdss.org/dr12/spectro/galaxy\_mpajhu/}} \citep{2003MNRAS.341...33K}. Galaxies whose SFR are not available in the MPA-JHU catalog are not shown in the figure. We can see that our \oiii$\lambda$4363 selected galaxies are located in the upper star forming region, and their total M$_*$ are lower than the rest of the normal SFGs. The median mass (log (M$_*$/M$_\odot$)) of our \oiii$\lambda$4363 selected galaxies is 9.25 and the median mass (log (M$_*$/M$_\odot$)) of SFGs in G18 is 9.68. However, we have noticed that some of our sample are located too low on the M$_*-$SFR diagram, indicating that these galaxies may not be star-forming. We have checked these galaxies and find that there are indeed local star-forming regions in these low-SFR galaxies. This indicates that a galaxy with a lower global SFR may also has more intense star formation locally.

\begin{figure*}[ht!]
	\includegraphics[width=1\textwidth]{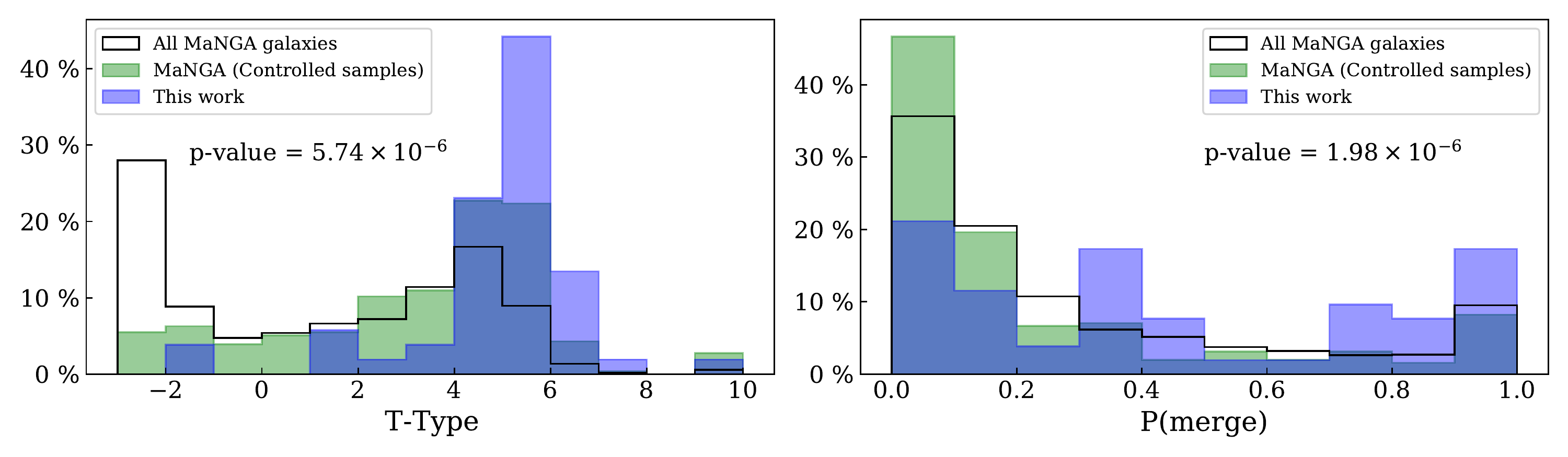}
	\caption{The normalized morphology properties distribution. Morphology data are from the MaNGA Deep Learning DR15 Morphology catalog \citep[DL;][]{2018MNRAS.476.3661D}. Blue filled histograms are the final \oiii$\lambda$4363 sample in this work. Black hollow histograms are the distribution of all galaxies in the DL catalog. Green filled histograms are the controlled sample of galaxies whose M$_*$ and SFR distributions are similar to the distribution of our final sample. Each vertical axis is the percentage of the total sample occupied by each bin. The $p$-values of Kolmogorov-Smirnov test between our final sample and controlled sample is also shown in each panel. \label{fig:overall_galaxies_properties}}
\end{figure*}

Subsequently, we match the MaNGA Deep Learning DR15 Morphology catalog \citep[DL;][]{2018MNRAS.476.3661D} to extract their morphological parameters, the T-Type value and the probability of merger, which have been widely used in previous works \citep[e.g.,][]{10.1093/mnras/stu2333,10.1093/mnras/sty3135,Chen_2020}. This catalog is trained by Deep Learning algorithms using Convolutional Neural Networks based on the catalog of \cite{2010ApJS..186..427N} and color images. Due to the particularity of our sample in M$_*$ and SFR, which may cause a systematic offset in morphological parameters, we select galaxies with the same M$_*$ and SFR distributions as our sample from the MaNGA survey to construct a controlled sample. The distributions of morphology are shown in Figure \ref{fig:overall_galaxies_properties}. Here we regard galaxies with a merger probability greater than 0.7 as merger. We found that 18 galaxies in our 52 sample galaxies are merging, accounting for 34.6\%, which is higher than the 14.8\% of all MaNGA galaxies and 13.0\% of controlled sample. We also perform a Kolmogorov-Smirnov test between our sample and the controlled sample and get the $p$-value of T-Type $\sim$ 10$^{-6}$, and the $p$-value of P(merge) $\sim$ 10$^{-10}$. Table \ref{tab:galaxies_info} show the picture and information of the 30 galaxies with the most spaxels in our final, which occupy the main part ($\sim$84\%) of our all spaxels. From their pictures, we can see that they are all blue in color and irregular in morphology, and some of them seem to be interaction. These observational features are consistent with the morphological parameters from the DL catalog. More details will be discussed in Section \ref{subsec:merger}. Therefore, we can conclude that the galaxies we selected not only have lower M$_*$ and higher SFR, but also later type and higher merger ratio in morphology, and this is not because of their low M$_*$ and high SFR.

We also check the locations of our selected spaxels in these sample galaxies. We find that most spaxels are located within a single star forming region. Regions of most galaxies (42 galaxies) are located in the periphery of the galaxy. Regions of 8 galaxies are in the center of the galaxy. 2 are distributed in the entire galaxy, and the remaining two can not be distinguished because their host galaxies have lower projected axis ratio.

\begin{figure}
    \centering
    \includegraphics[width=1\columnwidth]{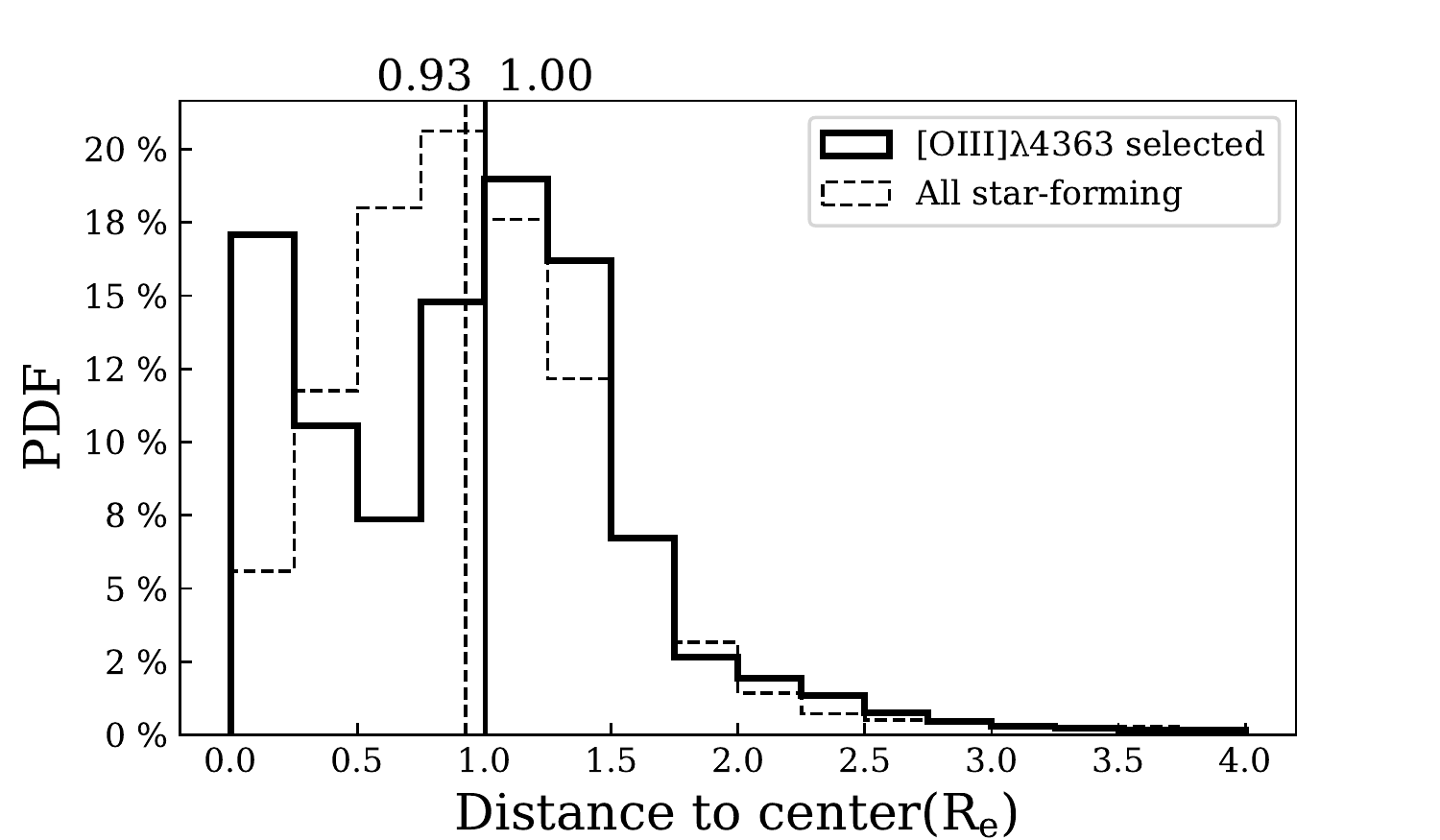}
    \caption{The normalized distribution of the distance to the center of galaxy. Distance is relative to R$_{\rm e}$, corrected by projection. Solid border histograms represent the sample we selected by \oiii$\lambda$4363. Dashed border histograms represent all star-forming spaxels in our sample galaxies. The vertical lines respectively indicate the median of the corresponding border style.}
    \label{fig:position_distribution}
\end{figure}

Here we compare the location distribution of our \oiii$\lambda$4363 selected spaxels in our 52 galaxies and the star-forming spaxels in the same galaxy (Figure \ref{fig:position_distribution}). The distances to the center are in units of effective radius(R$_{\rm e}$) and have already been corrected by $b/a$. It is found that there is no obvious deviation between the location distribution of our regions and the overall star-forming regions.

\subsection{Local Properties and Scaling Relations} \label{subsec:local_result}

\begin{figure}[ht!]
	\centering
	\includegraphics[width=1\columnwidth]{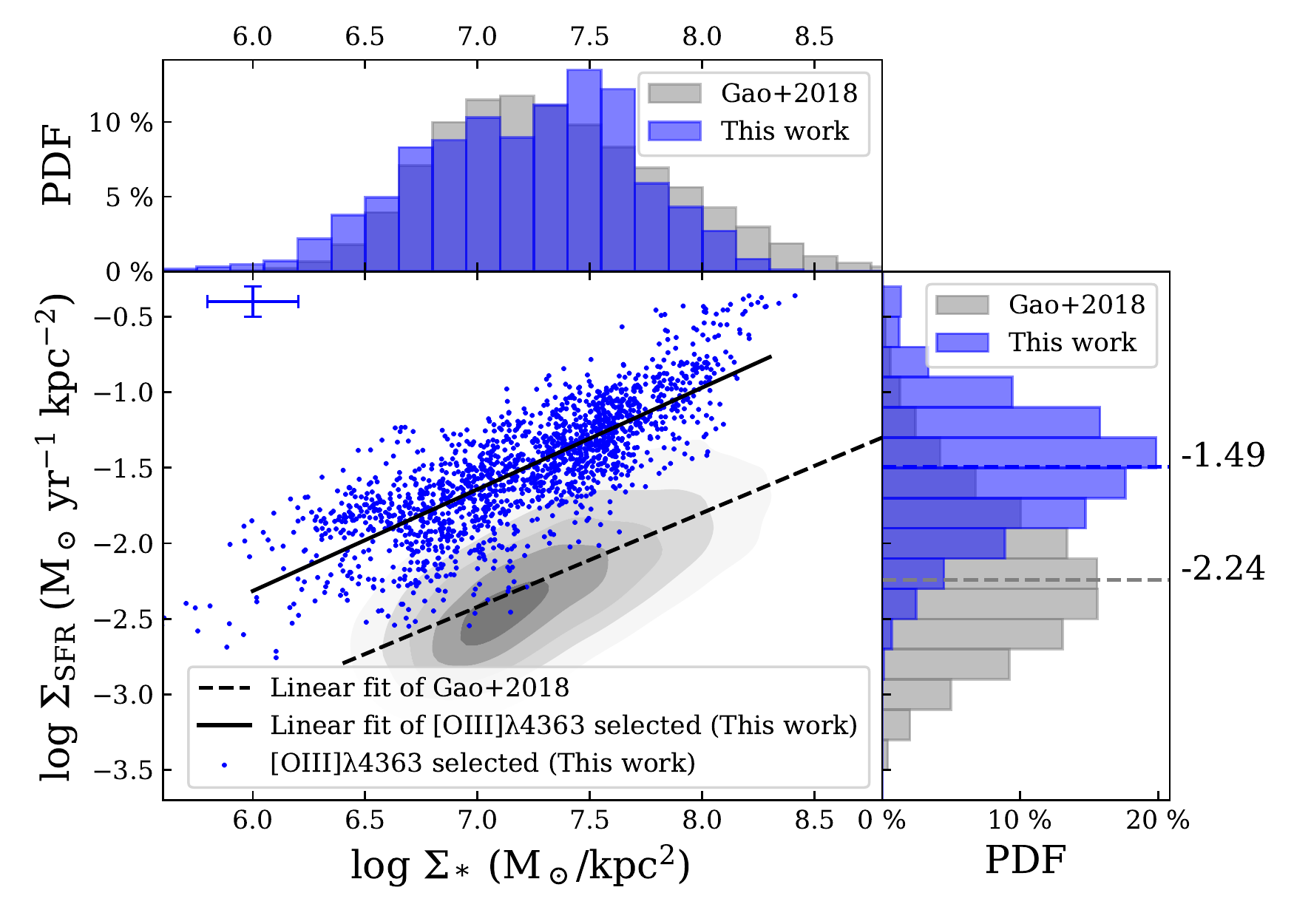}
	\caption{Local SFMS: blue represents our \oiii$\lambda$4363 selected spaxels, and gray represents star forming spaxels from G18. Histogram: the normalized distribution of the values corresponding to the coordinate axis. The dotted line indicates the median of the distribution of the corresponding color. Scatter diagram: the local main-sequence relation and bottom panel is the $\Sigma_*-Z$ relation. Blue dots are our sample, and the gray contour lines represent the distribution of 85\% of normal star forming spaxels in G18, and each contour encloses 15\%. Black solid lines are the linear fit of our sample. The dashed line is the linear fit of G18's local SFMS. The error bars of x and y axis are the uncertainty of $\Sigma_*$ and $\Sigma_{\rm SFR}$ described in Section \ref{subsec:uncertainty} with an error of $\sim$0.20 dex and $\sim$0.10 dex, respectively. \label{fig:sigma_sfr_relation}}
\end{figure}

\begin{figure}[ht!]
	\centering
	\includegraphics[width=1\columnwidth]{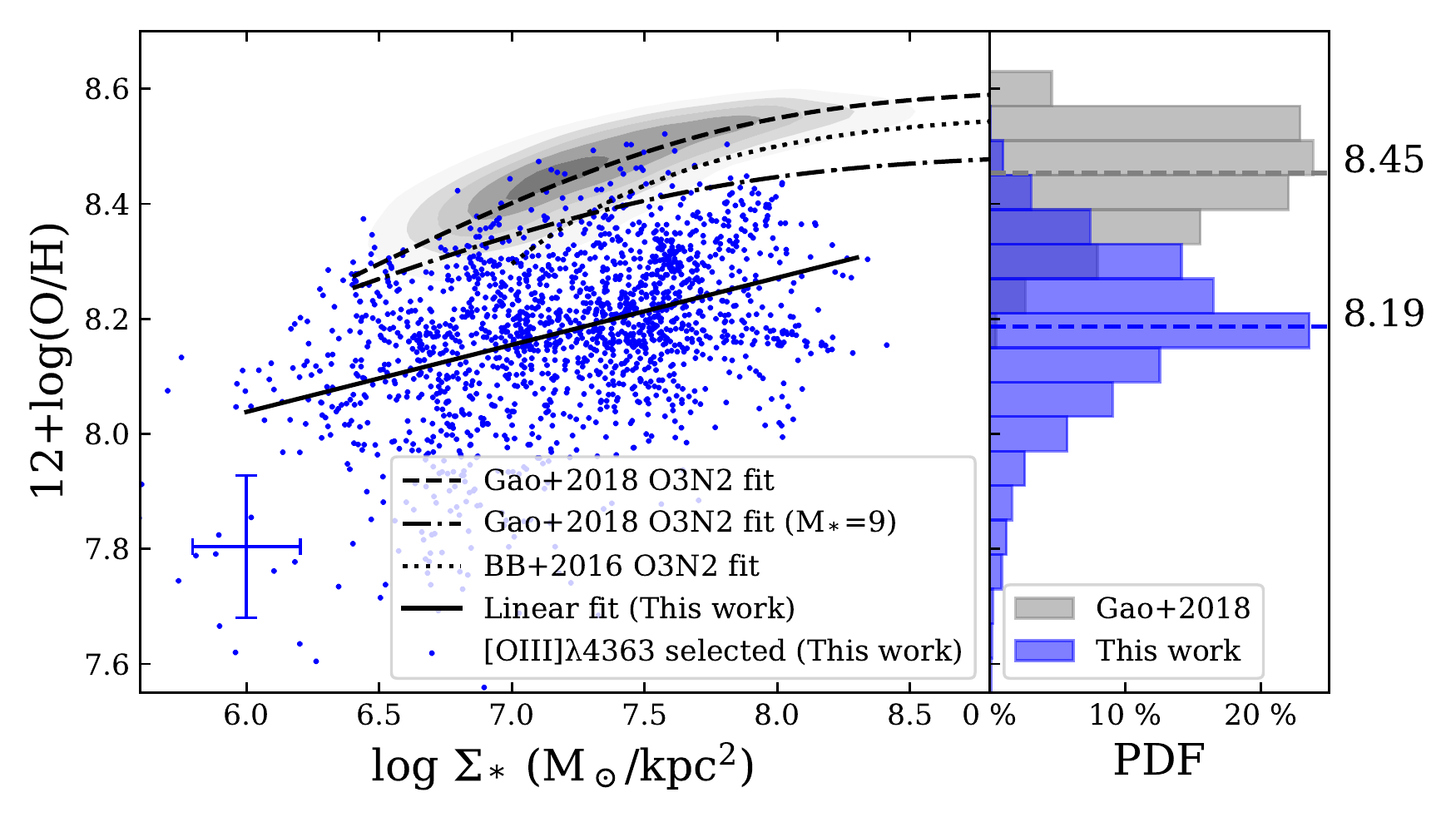}
	\caption{Local MZR: legends are the same as Figure \ref{fig:sigma_sfr_relation} but the dashed curve is the fit of G18's $\Sigma_*-Z$ relation, and the dotted dashed curve is the fit of G18's M$_*-\Sigma_*-Z$ relation when M$_*$=9. The dotted curve is from \cite{2016MNRAS.463.2513B}. The error bars of x and y axis are the uncertainty of $\Sigma_*$ and metallicity described in Section \ref{subsec:uncertainty} with an error of $\sim$0.20 dex and $\sim$0.12 dex, respectively. \label{fig:sigma_z_relation}}
\end{figure}

Since MaNGA has spatially resolved IFU spectra, the main focus of this work is on the local properties of these galaxies, such as $\Sigma_*$, $\Sigma_{\rm SFR}$, local metallicity and their scaling relationship. The $\Sigma_*-\Sigma_{\rm SFR}$ relationship is plotted in Figure \ref{fig:sigma_sfr_relation}, and $\Sigma_*-Z$ is plotted in Figure \ref{fig:sigma_z_relation}. The edges of the diagram show the normalized distributions of the corresponding parameters. The scatter panel shows the local SFMS: $\Sigma_*-\Sigma_{\rm SFR}$ relation for our \oiii$\lambda$4363 selected spaxels. We perform a linear fit to our spaxels and find that

\begin{equation}
{\log \rm \Sigma_{SFR} = (0.67\pm0.01) \log \Sigma_* - (6.35\pm0.09)}.
\end{equation}
The correlation coefficient is 0.80, and the residual standard deviation is 0.24. The uncertainties of slope and intercept are obtained by 1000-times bootstrap method, and following fittings are also the same. The units of $\Sigma_*$ and $\Sigma_{\rm SFR}$ are M$_\odot$ kpc$^{-2}$ and  M$_\odot$ yr$^{-1}$ kpc$^{-2}$ respectively. Similar fit is done for the star forming spaxels in G18 and results in

\begin{equation}
{\log \rm \Sigma_{SFR} = (0.61\pm0.00) \log \Sigma_* - (6.71\pm0.03)}.
\end{equation}
The correlation coefficient is 0.61, and the residual standard deviation is 0.39. We can see that our sample have higher $\Sigma_{\rm SFR}$ than G18's sample at a given $\Sigma_*$. There is also a significant gap in the median of log $\Sigma_{\rm SFR}$. Ours is -1.49 while G18's is -2.24. Only 10 spaxels (0.56\%) of our sample are within 1$\sigma$ (68\%) dispersion of the local SFMS of G18. These results indicate that the SFR of our sample is higher than that of the star formation sequence.

Figure \ref{fig:sigma_z_relation} shows the local MZR ($\Sigma_*-Z$ relation) of our sample and normal star forming sample based on O3N2 method in G18. We note that the O3N2 calibration adopted by G18 is the ONS-based one from \cite[][here after M13]{2013A&A...559A.114M}. This may cause a systematic deviation from the $T_{\rm e}$ method. We thus transform all metallicities data in G18 into the $T_{\rm e}$-based calibration by the following equation:

\begin{equation}\label{eq:o3n2_transform}
\begin{split}
\rm 12+log(O/H)_{O3N2,T_e} = 8.533\\
\rm -\frac{8.505-12+log(O/H)_{O3N2,ONS}}{0.221} \times 0.214
\end{split}
\end{equation}

For our sample, it is also found that the metallicity increases with increasing $\Sigma_*$, but the dispersion is large. We find that 221 spaxels (13.0\%) of our sample are within 1$\sigma$ dispersion of the local MZR of G18. We simply perform a linear fit to our spaxels and find that

\begin{equation} \label{eq:sigma_z_fit}
{\rm 12+\log{(O/H)} = (0.12\pm0.01) \log \Sigma_* + (7.34\pm0.05)},
\end{equation}
with a correlation coefficient of 0.41, and a residual standard deviation of 0.13.

We plot the best fit $\Sigma_*-Z$ relation of G18's radial bin sample. Taking into account the weak influence of the total M$_*$ on the local metallicity \citep{2016MNRAS.463.2513B,2018ApJ...868...89G} and the smaller M$_*$ of our galaxies (see Section \ref{subsec:global_result}), the best-fit M$_*-\Sigma_*-Z$ relation at M$_*=10^9$ M$_{\odot}$ is also given to eliminate the influence of the total stallar mass of galaxies.

In Figure \ref{fig:sigma_z_relation}, we can see that the $\Sigma_*-Z$ relation of our \oiii$\lambda$4363 selected spaxels is still systematically lower than that of the normal star forming spaxels in MaNGA when considering the total M$_*$. These spaxels with high $\Sigma_*$ have anomalously low metallicity (ALM), which are similar to the ALM regions mentioned in \cite{2019ApJ...872..144H}. More details will be discussed in Section \ref{subsec:merger}.

\begin{figure*}[ht!]
	\includegraphics[width=1\textwidth]{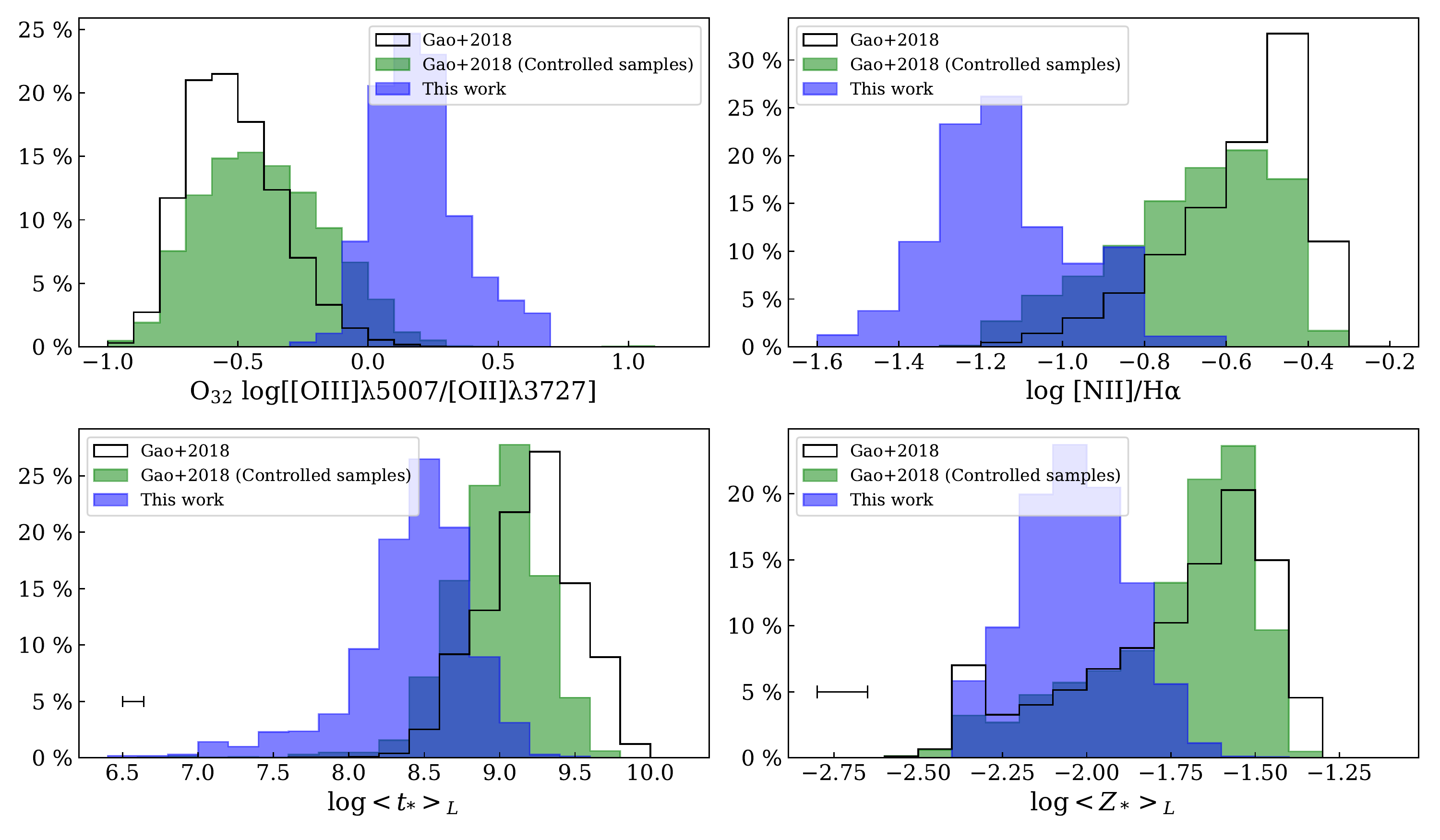}
	\caption{Normalized local properties distribution. Blue filled histograms are the final \oiii$\lambda$4363 spaxels in this work. Black hollow histograms are the distribution of all star forming spaxels in G18. Green filled histograms are the controlled sample of spaxels in G18 whose $\Sigma_*$ and $\Sigma_{\rm SFR}$ distribution are similar to the distribution of our sample. Each vertical axis is the percentage of the total sample occupied by each bin. Top 2 histograms: distributions of O$_{32}$ and N2 diagnostics. Bottom 2 histograms: distributions of light-weighted stellar properties. The unit of $t_*$ is year, and $Z_*=Z/Z_\odot$. The error bars are uncertainties from STARLIGHT when the signal-to-noise ratio of continuum is 5. \label{fig:overall_spaxels_properties}}
\end{figure*}

We next investigate and compare the local ionized gas properties and stellar properties through emission lines and continuum of spaxels. 

The upper two panels of in Figure \ref{fig:overall_spaxels_properties} show distributions of two emission line ratios O$_{32}$=log(\oiii($\lambda$4959+$\lambda$5007)/\oii$\lambda$3727) and N2=log(\nii$\lambda$6583/H$\alpha$), while the lower ones are the distributions of the luminosity-weighted $t_*$ and $Z_*$ obtained by STARLIGHT fitting. Among them, O$_{32}$ can roughly characterize ionization parameters \citep{2019ARA&A..57..511K}, and N2 can roughly characterize their metallicities \citep{2004MNRAS.348L..59P}. The unit of $t_*$ is year, and stellar metallicity $Z_*=Z/Z_\odot$. When the S/N of continuum is 5, the uncertainties of log$\langle t_*\rangle_L$ and log$\langle Z_*\rangle_L$ are 0.14 dex and 0.15 dex respectively \citep{2005MNRAS.358..363C}, which are displayed by error bars.

Since the SFR of our sample is relatively high overall, and we have noticed the possible local Fundamental Metallicity Relation \citep[FMR;][]{2010MNRAS.408.2115M} which means that regions with higher SFRs tend to have lower metallicities at fixed $\Sigma_*$. In order to verify whether our low metallicities are the result of high $\Sigma_{\rm SFR}$ through FMR, here we also need to control the sample likes Section \ref{subsec:global_result}. Green histograms in Figure \ref{fig:overall_spaxels_properties} show the distributions of the corresponding properties for the sub-sample randomly selected from G18, containing 2499 spaxels, whose distribution on the $\rm \Sigma_*-\Sigma_{SFR}$ plane is the same as that of our final \oiii$\lambda$4363 sample. Thus we can eliminate the potential impact of FMR.

The distribution of data shows that the ionization parameters, gas phase metallicity, $t_*$ and $Z_*$ of our sample are obviously different from those of all sample and the controlled sample in G18. There is only a small difference between the controlled sample and all sample. Therefore, we can conclude that the spaxels we selected not only have higher $\rm \Sigma_{SFR}$, but also higher ionization parameter, lower gas phase and stellar metallicity and younger stars, and this is not because of their high $\rm \Sigma_{SFR}$ under the impact of FMR.

\subsection{Nitrogen abundances} \label{subsec:no_result}
\begin{figure}[ht!]
	\centering
	\includegraphics[width=1\columnwidth]{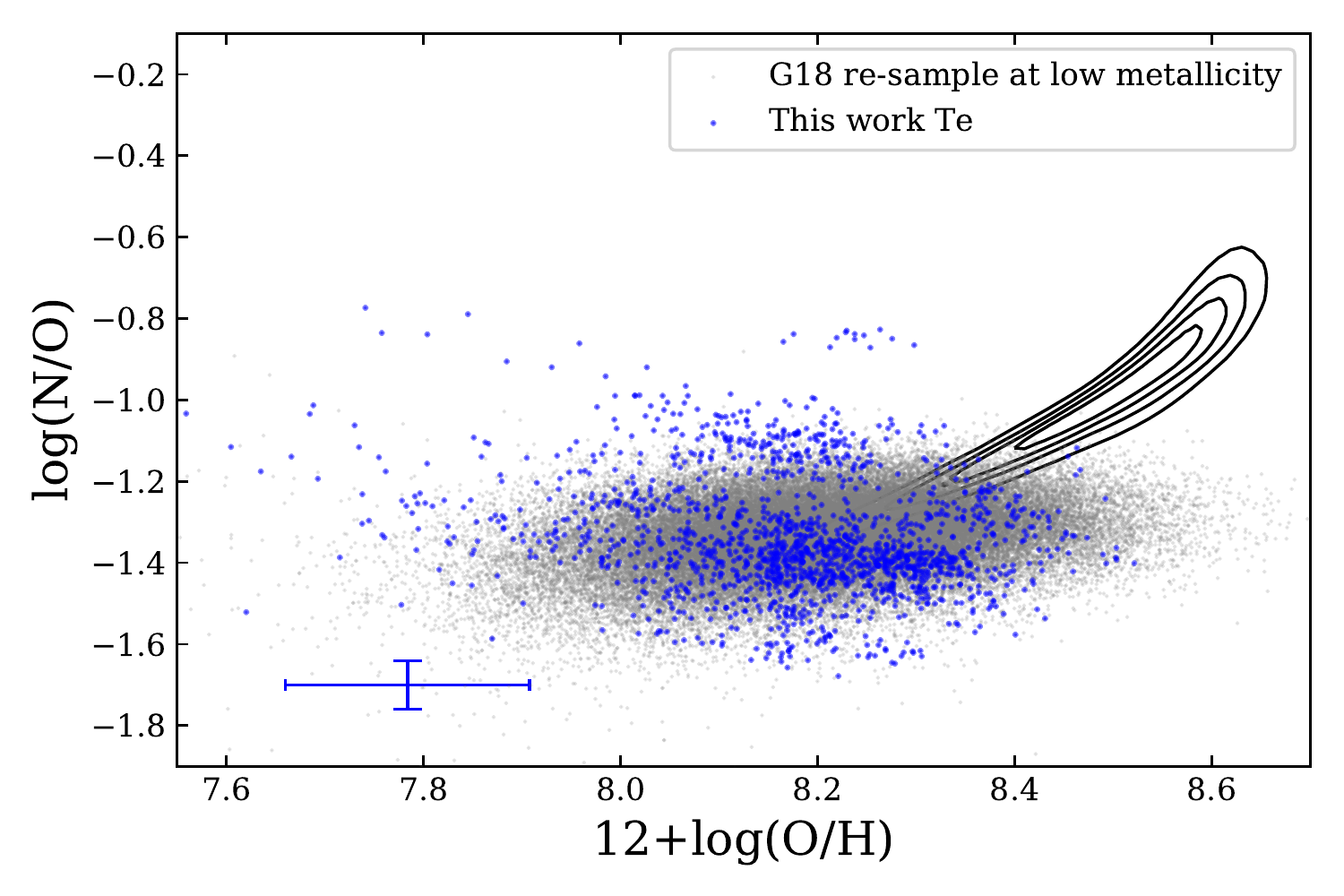}
	\caption{The 12+log(O/H) $-$ log(N/O) relation. The blue dots represent the T$_{\rm e}-$based result of this work. The black solid contours trace the number density distribution of G18 spaxels, whose O and N abundances are both derived from PG16, and the gray dots correspond to re-sampling 12+log(O/H) $<$ 8.3 data points of G18 by randomly adding Gaussian noise with the same standard deviation as blue dots. The error bar in the lower left corner represents the typical uncertainty of our sample on the two axes. \label{fig:no_relation}}
\end{figure}
{\bf Metals are direct products of stellar nucleosynthesis, elements originate from different nucleosynthetic can probe the star formation history and chemical enrichment in galaxies. Here we calculate the abundance rations between nitrogen (mostly produced via primary nucleosynthesis in low metallicity regions, as well as secondary nucleosynthesis of low-mass and intermediate-mass stars from higher metallicity regions) and oxygen (mostly produced by massive stars on short time-scales). The N/O  probes the star formation time scale, and N/O $-$ O/H diagram can also provides a test of the validity of  our Te-based abundances.The calculation method used is also from \cite{2010ApJ...720.1738P}. Since the \nii$\lambda$5755 line is too weak to be detected, we use t$_2$ estimated from t$_3$ instead of t$_{\rm 2,N}$ to derive the nitrogen abundance. The 12+log(O/H) $-$ log(N/O) relation is shown in Figure \ref{fig:no_relation}. The uncertainty of log(N/O) is $\sim$0.06 dex, which is obtained using the same process as Section \ref{subsec:uncertainty}. From Figure \ref{fig:no_relation} we can find that our samples are distributed at the flat low end of the relationship, and compared with the G18 sequence, there is much greater dispersion and no positive correlation. This is because the nitrogen at the low metallicity end is mainly produced by primary nucleosynthesis, which produces similar amounts of N and O. As shown by the distribution of randomly re-sampled grey dots of G18 at low metallicity , the large dispersion of our metallicity measurement is a nature result of the uncertainty.}

\section{Discussion} \label{sec:discussion}
\subsection{Direct Comparison of Strong Line Method} \label{subsec:strong_line_cmp}
In this work, we directly compare the metallicity of our \oiii$\lambda$4363 selected sample with that of G18. Although we have corrected the systematic error of the O3N2-ONS metallicity calculation, after all, the O3N2 strong line method is particularly sensitive to ionization parameters \citep{2004MNRAS.348L..59P}, and our sample has an excessively high ionization parameter that deviates from the normal, so it is necessary to use the same strong line method to compare their metallicity again.

\begin{figure}[ht!]
	\centering
	\includegraphics[width=1\columnwidth]{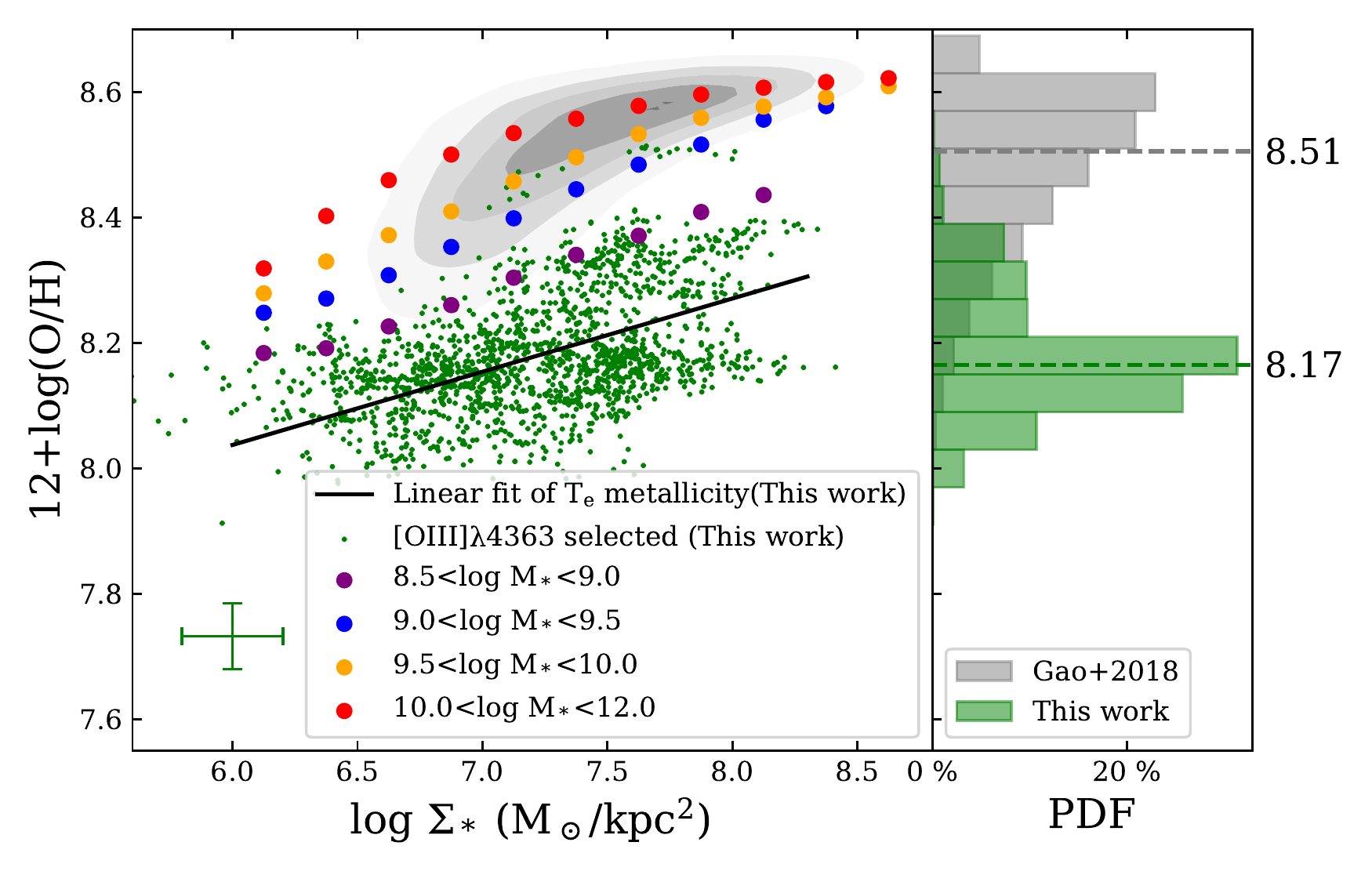}
	\caption{The legends are the same as Figure \ref{fig:sigma_z_relation}, but green replaces blue. The black solid line still represents the fitting result given by the electronic temperature method data (Equation \ref{eq:sigma_z_fit}), while green dots and histograms represent the result of the strong line method. The colored solid circles represent the median of metallicity in each M$_*$ and $\Sigma_*$ bin for interpolation. \label{fig:sigma_z_o3n2_relation}}
\end{figure}

Here we use PG16-R-calibrated metallicity, because the applicable metallicity range of M13 calibration is too narrow to include all our spaxels, while the range of PG16 calibration is sufficient to include our sample and G18 sample. The uncertainty of metallicity is also obtained using the same process as Section \ref{subsec:uncertainty}. The final uncertainty of PG16-R-calibrated metallicity is $\sim$0.05 dex. For the G18 sample, We re-calculate their metallicity using the R calibration of PG16 to keep consistency.

The new $\Sigma_*-Z$ relationship is shown in Figure \ref{fig:sigma_z_o3n2_relation}. In addition, we also performed an interpolation fitting on PG16-R-calibrated metallicities of G18 to get a PG16-R-calibrated M$_*-\Sigma_*-Z$ relation. The interpolation process is the same as \cite{2019ApJ...872..144H}, so that we can get the expected fitting value of PG16 metallicity ($Z_{\rm fit,PG16}$) under a given M$_*$ and $\Sigma_*$. The result of interpolation is shown in the colored solid circles in the figure, and the standard deviation between our interpolation result and the observation of G18 is 0.07 dex. Since the statistical uncertainties arising from the emission line measurements are much smaller than the systematic uncertainty (0.049 dex) of the calibration, the error bar of the metallicity is replaced by the systematic uncertainty of the adopted calibration. We can see that after switching to the strong line method, the $\Sigma_*-Z$ relation still exists, and it moves down. The median metallicity of the $T_{\rm e}$ method is 8.22, while the PG16 method is 8.17. Under this method, only 64 spaxels ($\sim$3.8\%) are within the 1$\sigma$ dispersion of the local MZR of G18. We can confirm the overall low metallicities of our sample and the existence of the local $\Sigma_*-Z$ relationship.

\subsection{Comparison with All Star Forming Regions in Sample Galaxies and Other Galaxies with Similar M$_*$ Distribution} \label{subsec:compare}

We have determined that our sample galaxies have low total stellar masses in Section \ref{subsec:global_result}, and their metallicities are lower than the best fit of G18's M$_*-\Sigma_*-Z$ relation at M$_*$=$10^9$M$_\odot$. However it is known that the metallicities of the galaxies of a given total stellar mass show a significant scatter, especially at low masses. We should discuss further and clarify whether the metallicity of the selected regions only are lower in comparison to the metallicities of other regions in the same galaxy, or the metallicities of galaxies as whole from their sample are lower in comparison to other galaxies.

\begin{figure}[ht!]
	\centering
	\includegraphics[width=1\columnwidth]{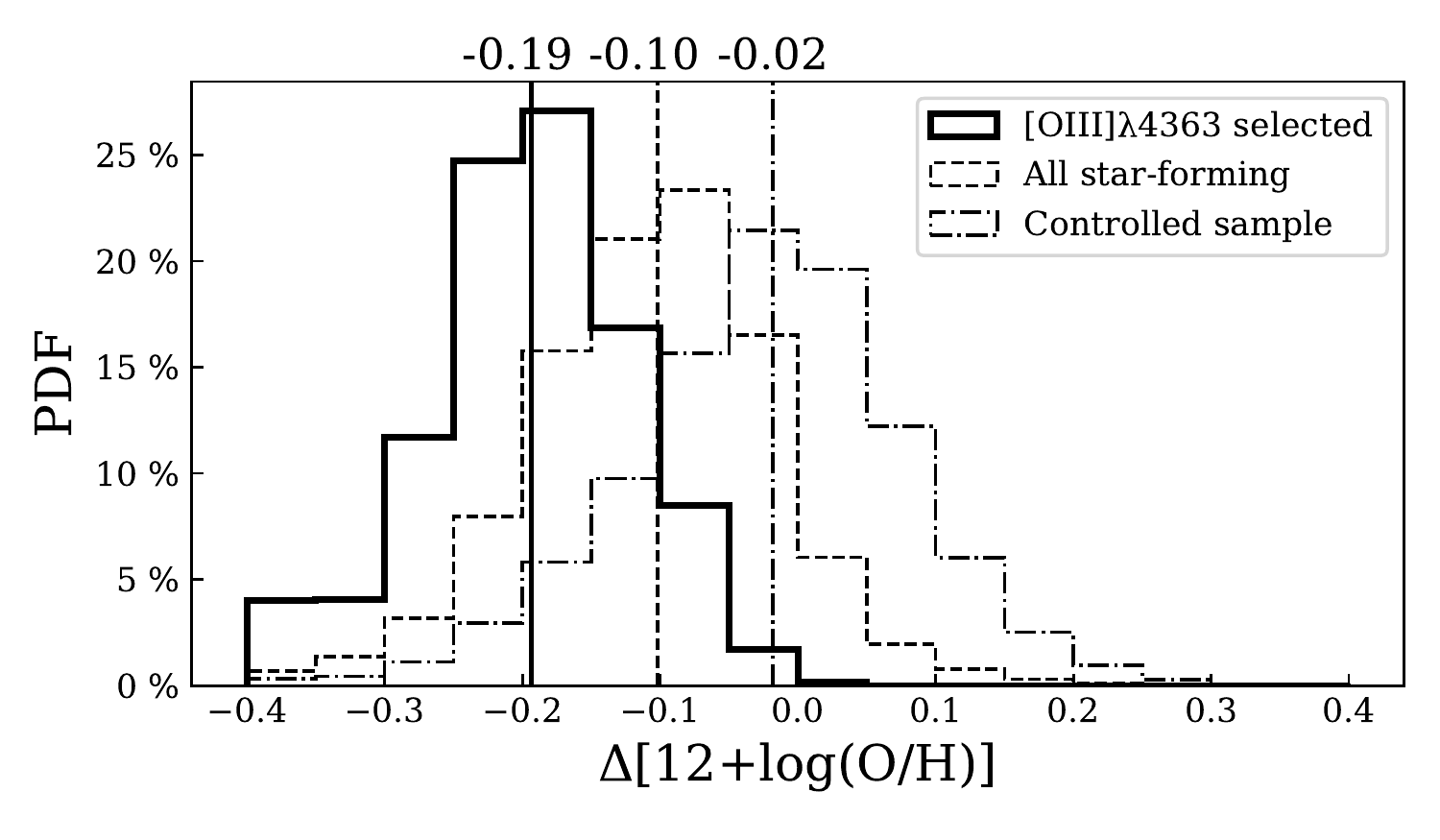}
	\caption{The normalized distribution of metallicity offset ($\Delta$[12+log(O/H)]=$Z_{\rm fit,PG16}(\Sigma_*$,M$_*)$-$Z_{\rm obs}$). All metallicities are obtained from PG16. Solid border histograms represent the sample we selected by \oiii$\lambda$4363. Dashed border histograms represent all star-forming spaxels in our sample galaxies. Dash-dot border histograms represent the control sample with similar mass distributions to our sample galaxies. The vertical lines respectively indicate the median of the corresponding border style. \label{fig:del_metal}}
\end{figure}

Like Section \ref{subsec:global_result}, we randomly selected 162 galaxies with similar mass distributions to our sample galaxies from MaNGA to make a control sample, and the O3N2 method (PG16) is also used to calculate their metallicities. In order to verify whether the metallicity is deviated from the normal star-forming sequence, we compare the metallicity of these samples with the interpolated M$_*-\Sigma_*-Z$ relation of G18 ($Z_{\rm fit,PG16}(\Sigma_*$, M$_*)$) in Section \ref{subsec:strong_line_cmp}, defining the metallicity offset, $\Delta$[12+log(O/H)]=$Z_{\rm fit,PG16}(\Sigma_*$,M$_*)$-$Z_{\rm obs}$. The results are shown in Figure \ref{fig:del_metal}. The median of \oiii$\lambda$4363 selected spaxels is -0.19, all star-forming spaxels is -0.10, and control sample is -0.02. This means that not only the metallicity of our \oiii$\lambda$4363 selected sample are lower than other star-forming regions in their host galaxies, but also the metallicities of these host galaxies are also entirely lower than other galaxies with similar M$_*$. Low metallicity is not only a local characteristic, but also a characteristic of the galaxy as a whole. Combined with the above results in \ref{subsec:global_result}, it shows that the appearance of these metal-poor regions is related to the particularity of the entire galaxy. We should notice that the selection effects caused by flux limits of  auroral emission lines may contribute to those results, we leave this consideration for future work when larger sample become avaliable.

\subsection{ALM Region and Merger} \label{subsec:merger}

In Section \ref{subsec:local_result}, Figures \ref{fig:sigma_z_relation} and \ref{fig:sigma_z_o3n2_relation}, we have found that our \oiii$\lambda$4363 selected metal-poor sample are not located at the low-metallicity end of the $\Sigma_*-Z$ relationship of normal star-forming regions. They seem to be ALM regions, selected by their significantly lower (about 0.14 dex) observed metallicities compared to the expected ones given $\Sigma_*$ and the $\Sigma_*-Z$ relation, in \cite{2019ApJ...872..144H}. These authors found that ALM regions tend to exist in less massive and disturbed galaxies. They also estimated the ages of these ALM regions to be a few hundred Myr. Here we use \oiii$\lambda$4363 emission line to select metal-poor sample to get the more accurate metallicities than strong line method. In our \oiii$\lambda$4363 selected metal-poor sample, the $t_*$ are also between 100 Myr and 1 Gyr ($\log\langle t_*\rangle \sim 8.0-9.0$). The distribution of P(merger) shown in the right panel of Figure \ref{fig:overall_galaxies_properties} also shows that our galaxies are likely to experience mergers or interactions compared to the controlled sample. These characteristics are highly consistent with the ALM regions selected by \cite{2019ApJ...872..144H}.

Since the P(merger) from DL catalog is the result of batch processing, it only gives a statistical merging probability. In fact, whether the merger is true requires further inspection, so we check our galaxies one by one. We review the images and spectra of each galaxy and its surrounding galaxies which may be merging objects. Our inspection finds that 11 of the 52 galaxies in our sample have nearby galaxies confirmed by spectral redshifts, of which 6 galaxies (plate-ifu=8566-3704, 8455-9101, 8241-6101, 8139-12702, 9881-9102, and 9194-12701) are within the field of view of the MaNGA IFU, and 3 galaxies (9509-3702, 8715-12704, and 8452-1902) are within 1\arcmin. In addition, the neighbors of 3 galaxies (8548-3702, 8553-12703, and 8325-12702) have no spectral redshift but are within the uncertainty of their photometric redshifts. Therefore, a total of 12 galaxies may be merging galaxies, accounting for 23.1\% of our sample. Therefore, the results of both the DL catalog and the spectral inspection confirm that our sample has a greater merger ratio than the normal SFGs. Merger and interaction can trigger gas inflow and star formation, and the inflow of metal-poor gas will dilute the metal-rich ISM and form new metal-poor stars. \citep{2000A&ARv..10....1K}, so we can conclude that the interaction is closely related to these metal-poor regions.

\section{Conclusion} \label{sec:conclusion}

In this work, we make use of the MaNGA data from the SDSS DR15 to investigate the $\Sigma_*-\Sigma_{\rm SFR}$ and $\Sigma_*-Z$ relationships. We select 1698 spaxels from 52 galaxies with reliable \oiii$\lambda$4363 observation. We calculate their local metallicities with the $T_{\rm e}$ method. We have got the following conclusions:
\begin{enumerate}
	\item There is a local $\Sigma_*-\Sigma_{\rm SFR}$ relationship and $\Sigma_*-Z$ relationship in the \oiii$\lambda$4363 emission line selected regions. Among them, in the $\Sigma_*-\Sigma_{\rm SFR}$ relationship, our SFR is higher overall (99.4\% are outside 1$\sigma$ of SFMS in G18), and in the $\Sigma_*-Z$ relationship, our metallicity is lower overall (80.5\% are outside 1$\sigma$), and this is not caused by the FMR.
	
	\item The regions we selected have young stars ($<$ 1 Gyr) and their stellar metallicities are also lower ($\sim$ 1\% Z$_\odot$). The $t_*$ derived by SED fitting is consistent with the estimated age of the ALM regions reported in \cite{2019ApJ...872..144H}.
	
	\item Most of the galaxies in the selected region are star-forming galaxies with low mass ($\sim$ 10$^9$ M$_\odot$). These galaxies are late in morphology and have a high merging ratio, reflecting that the metal-poor star-forming regions may be related to interactions. The inflow of metal-poor gas may dilute the interstellar medium and form new metal-poor stars in these galaxies during interaction.
\end{enumerate}

However, we do not find a direct evidence for the inflow of metal-poor gas and the relation between merger and gas inflow. How the interaction affects the metallicity of galaxies remains to be further studied.

\acknowledgments

This work is supported by the Strategic Priority Research Program of Chinese Academy of Sciences (No. XDB 41000000), the National Key R\&D Program of China (2017YFA0402600, 2017YFA0402702), and the NSFC grant (Nos. 11973038, 11973039 and 11421303). HXZ also thanks a support from the CAS Pioneer Hundred Talents Program.

Funding for SDSS-IV has been provided by the Alfred P. Sloan Foundation and Participating Institutions. Additional funding toward SDSS-IV has been provided by the US Department of Energy Office of Science. SDSS-IV acknowledges support and resources from the Centre for High-Performance Computing at the University of Utah. The SDSS website is www.sdss.org.

SDSS-IV is managed by the Astrophysical Research Consortium for the Participating Institutions of the SDSS Collaboration including the Brazilian Participation Group, the Carnegie Institution for Science, Carnegie Mellon University, the Chilean Participation Group, the French Participation Group, Harvard-Smithsonian Center for Astrophysics, Instituto de Astrofísica de Canarias, The Johns Hopkins University, Kavli Institute for the Physics and Mathematics of the Universe (IPMU)/University of Tokyo, Lawrence Berkeley National Laborato y, Leibniz Institut fur Astrophysik Potsdam (AIP), Max-Planck-Institut fur Astronomie (MPIA Heidelberg), Max-Planck-Institut fur Astrophysik (MPA Garching), Max-Planck-Institut fur Extraterrestrische Physik (MPE), National Astronomical Observatory of China, New Mexico State University, New York University, University of Notre Dame, Observatario Nacional/MCTI, The Ohio State University, Pennsylvania State University, Shanghai Astronomical Observatory, United Kingdom Participation Group, Universidad Nacional Autonoma de Mexico, University of Arizona, University of Colorado Boulder, University of Oxford, University of Portsmouth, University of Utah, University of Virginia, University of Washington, University of Wisconsin, Vanderbilt University, and Yale University.


\newpage
\bibliography{sample63}{}
\bibliographystyle{aasjournal}

\appendix
\section{Comparison of Our Results and DAP} \label{sec:comparison_dap}

\begin{figure*}[ht!]
	\centering
	\includegraphics[width=0.75\textwidth]{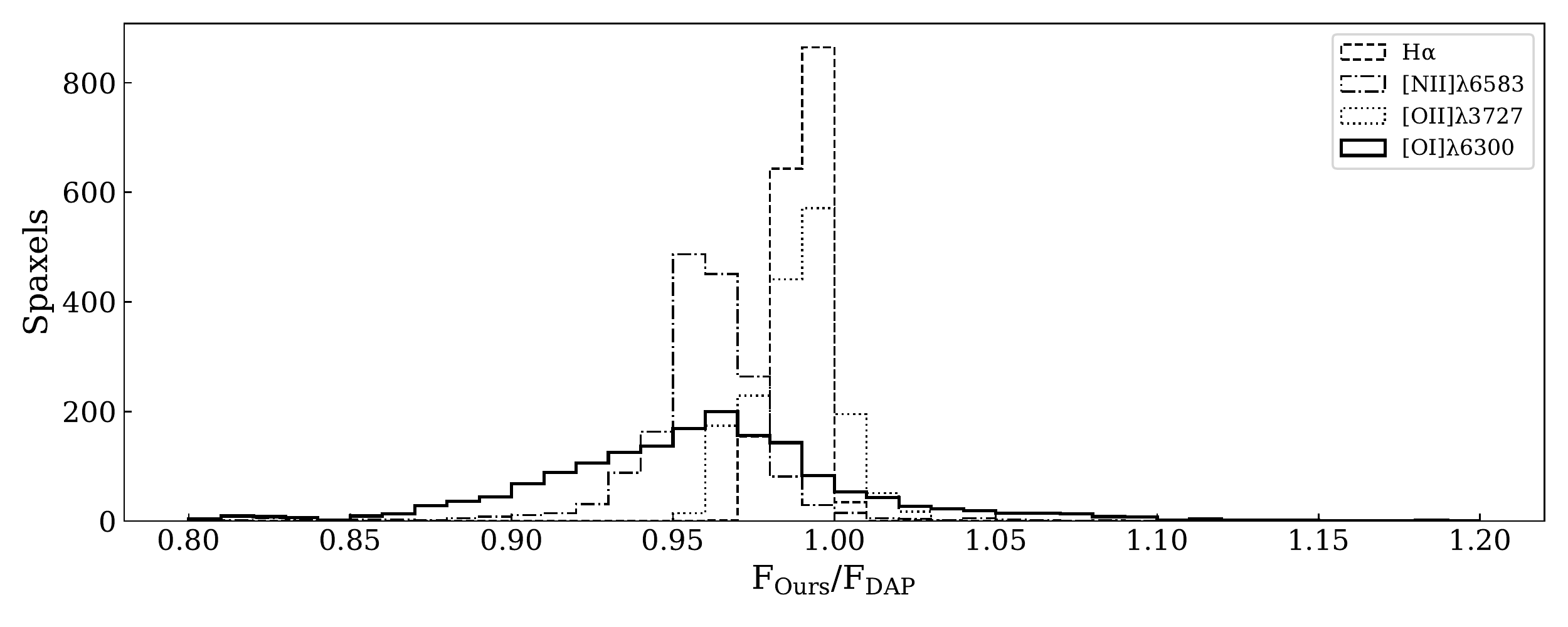}
	\caption{The distribution of the ratio of the flux values of some emission lines (\ha, \nii, \oii and \oi) we measured to the flux values provided by DAP at the same spaxel. \label{fig:comparison_dap_ratio}}
\end{figure*}

\begin{figure*}[ht!]
	\centering
	\includegraphics[width=0.8\textwidth]{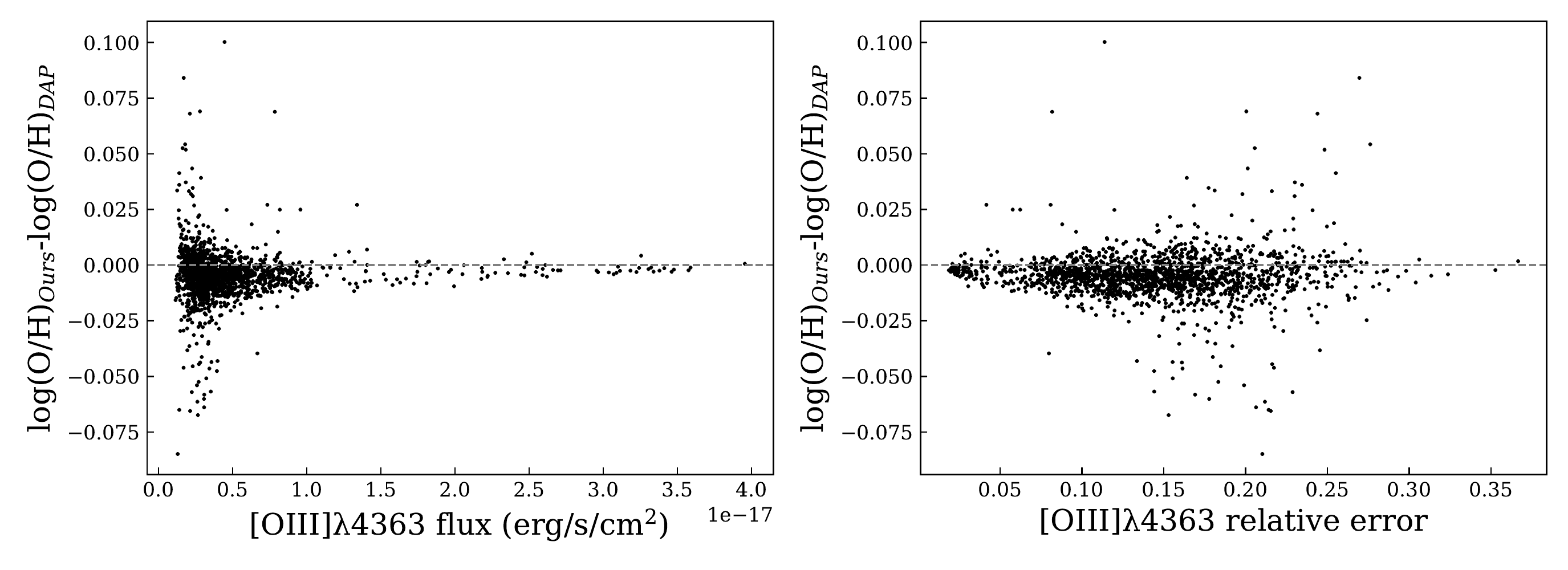}
	\caption{The difference between our metallicity and DAP-flux-based metallicity as the function of \oiii4363 flux and the relative error. \label{fig:metal_pg16_dap_d_vs_4363}}
\end{figure*}

\begin{figure*}[ht!]
	\centering
	\includegraphics[width=0.75\textwidth]{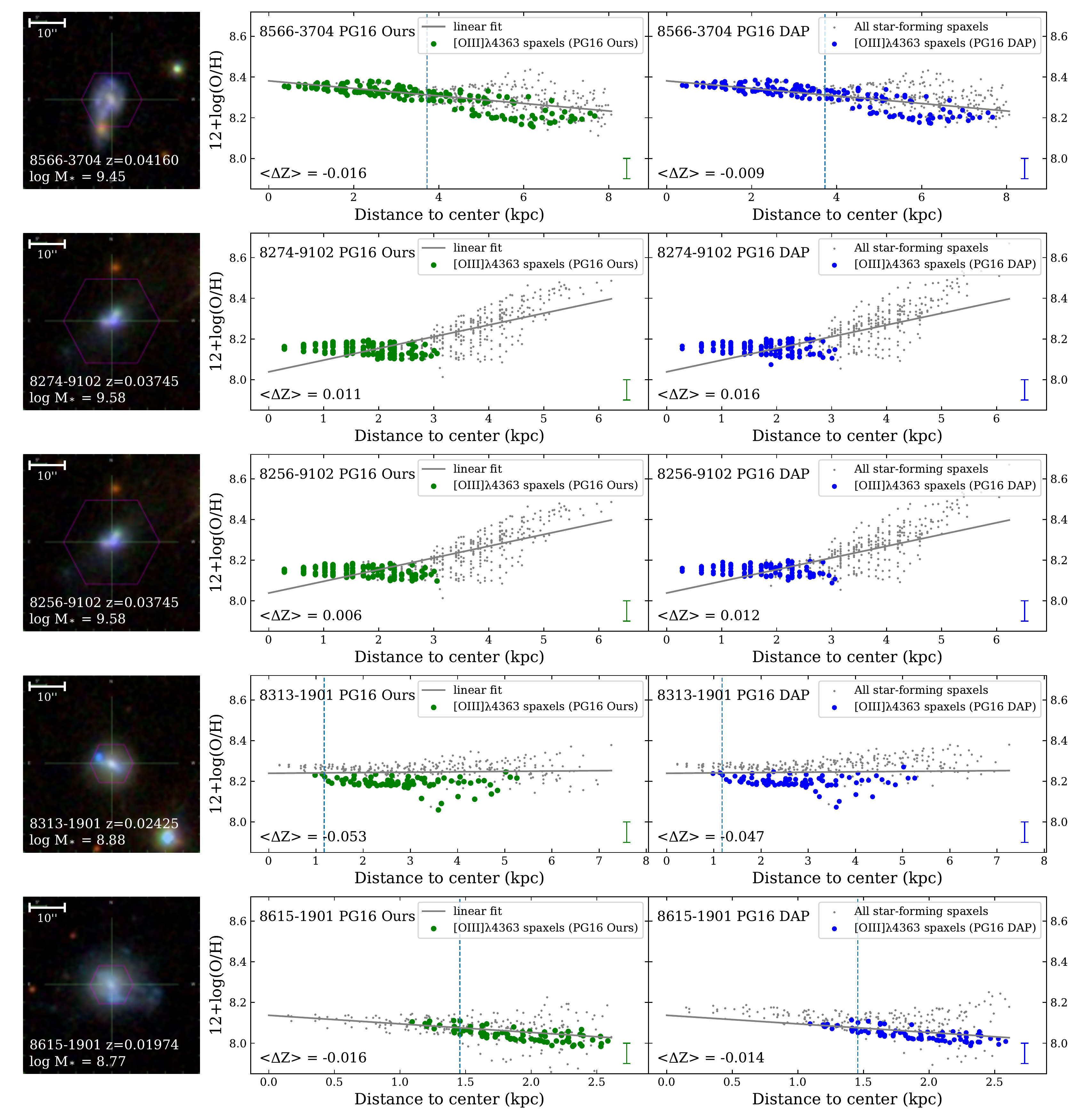}
	\caption{Same as Figure \ref{fig:metal_profile_0_5}, but the flux measurements of the right column are from DAP. \label{fig:metal_profile_dap}}
\end{figure*}

{\bf Here we provide a comparison between the results measured by us and provided by DAP. Figure \ref{fig:comparison_dap_ratio} shows the ratio of fluxes of some emission lines. The PG16-based differences between ours and DAP as functions of \oiii$\lambda$4363 flux and \oiii$\lambda$4363 relative error are shown in Figure \ref{fig:metal_pg16_dap_d_vs_4363}. The flux and error here are before the intrinsic dust extinction correction. Figure \ref{fig:metal_profile_dap} shows the metallicity profile measured by our fitting and DAP along radius of the first 4 galaxies which have already been shown in Figure \ref{fig:metal_profile_0_5}.

Since the emission lines we compared are all strong emission lines, it cannot prove that the measurement of the weaker emission lines (\oiii$\lambda$4363) is accurate. However, \oiii$\lambda$4363 is not measured by DAP, we can estimate its accuracy of fitting through \oi$\lambda$6300, whose flux is the same magnitude as \oiii$\lambda$4363's. The mean flux of \oiii$\lambda$4363 of our sample is $\sim$0.49$\times$10$^{-17}$erg/s/cm$^2$, and the mean flux of \oi$\lambda$6300 is $\sim$1.17$\times$10$^{-17}$erg/s/cm$^2$. The mean relative difference of \oi$\lambda$6300 measurement is only $\sim$6\%. Therefore, from Figure \ref{fig:comparison_dap_ratio} and \ref{fig:metal_profile_dap}, we can conclude that our measurement is reliable compared to DAP.

We also calculate the differences between two observations of the same galaxy (8274-9102 and 8256-9102), and they are shown in Figure \ref{fig:metal_te_8274_8256_d_vs_4363}.

\begin{figure*}[ht!]
	\centering
	\includegraphics[width=0.8\textwidth]{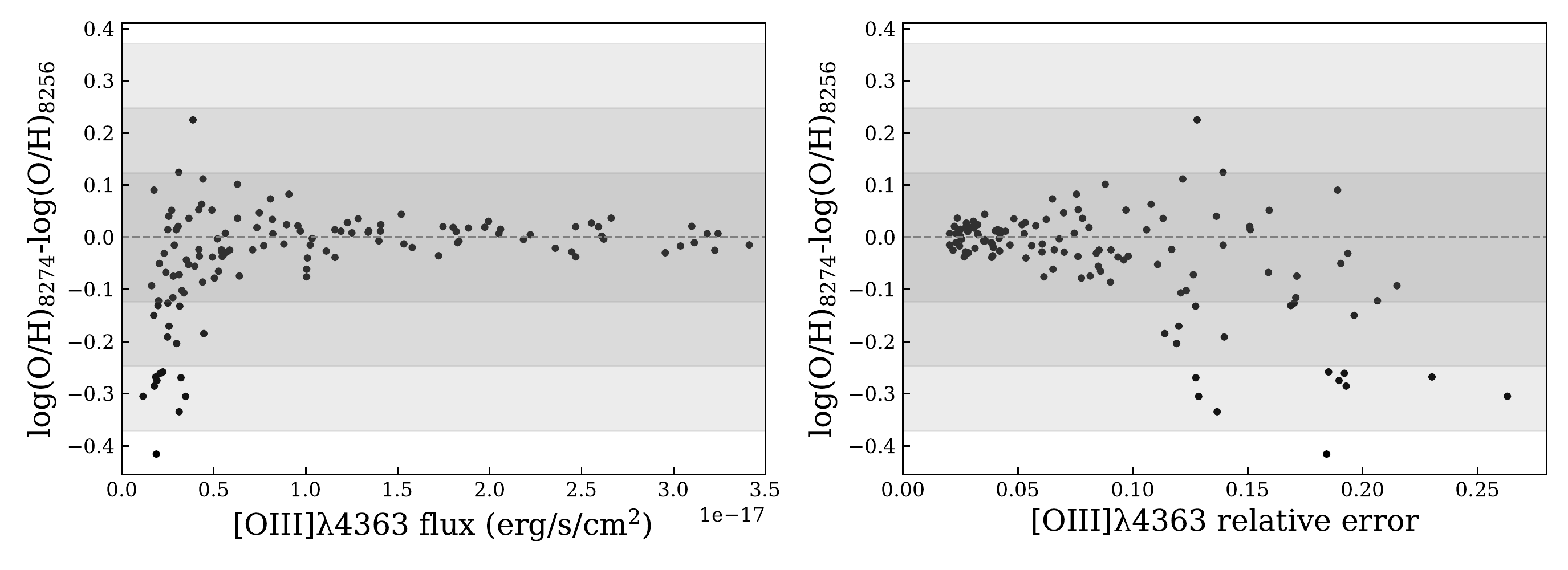}
	\caption{Same as Figure \ref{fig:metal_pg16_dap_d_vs_4363}, but the difference become the T$_{\rm e}$-based metallicity difference between 8274-9102 and 8256-9102. The typical metallicity uncertainty ($\sim$0.12dex) of one, two and three times is indicated by shading of different depths.  \label{fig:metal_te_8274_8256_d_vs_4363}}
\end{figure*}
}

\newpage

\section{Images of Other Sample Galaxies} \label{sec:other_galaxies}

\begin{figure*}[ht!]
	\centering
	\includegraphics[width=0.9\textwidth]{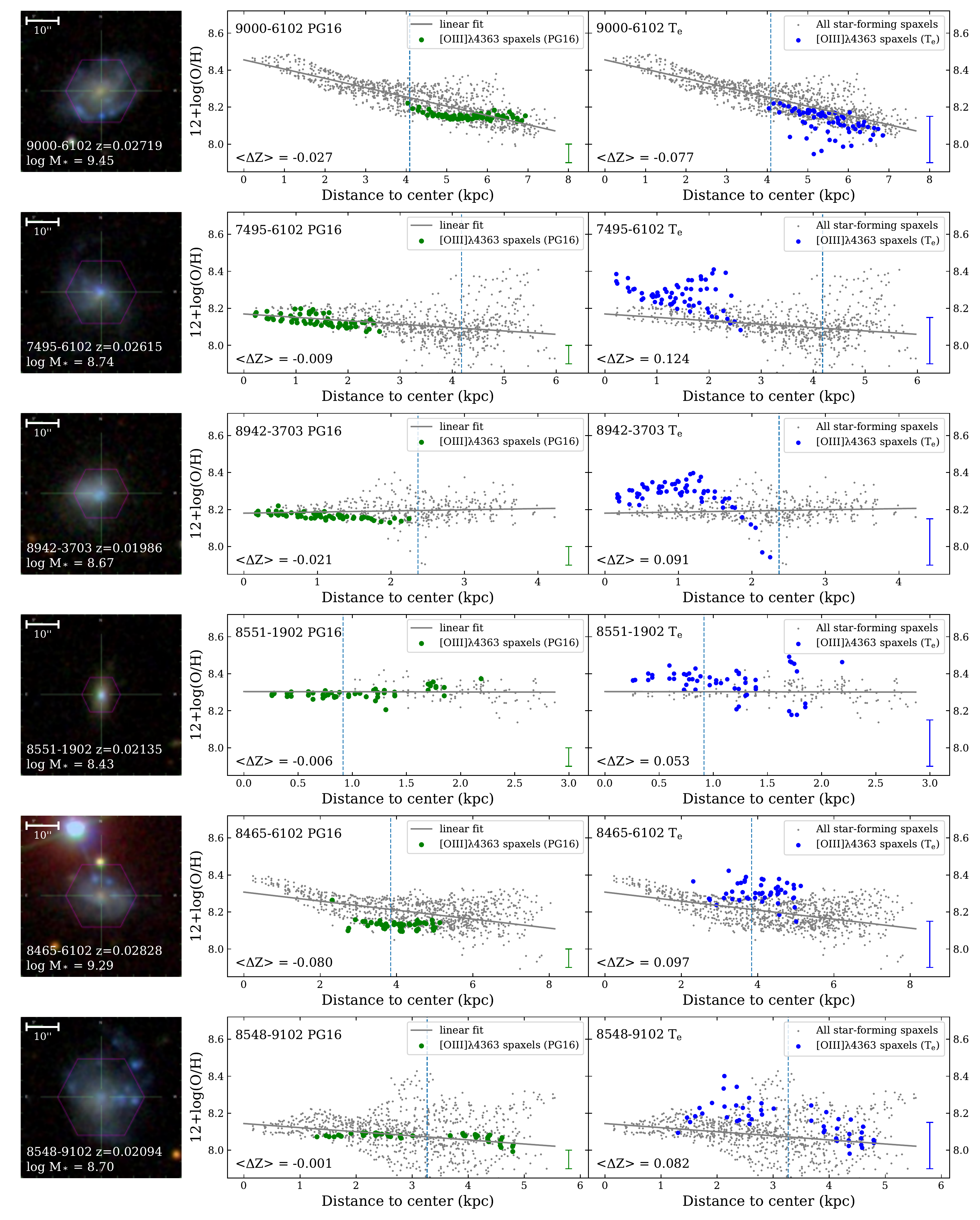}
	\caption{Same as Figure \ref{fig:metal_profile_0_5}, but the galaxies are ranked 5th to 10th in descending order of spaxels. \label{fig:metal_profile_5_11}}
\end{figure*}

\begin{figure*}[ht!]
	\centering
	\includegraphics[width=0.9\textwidth]{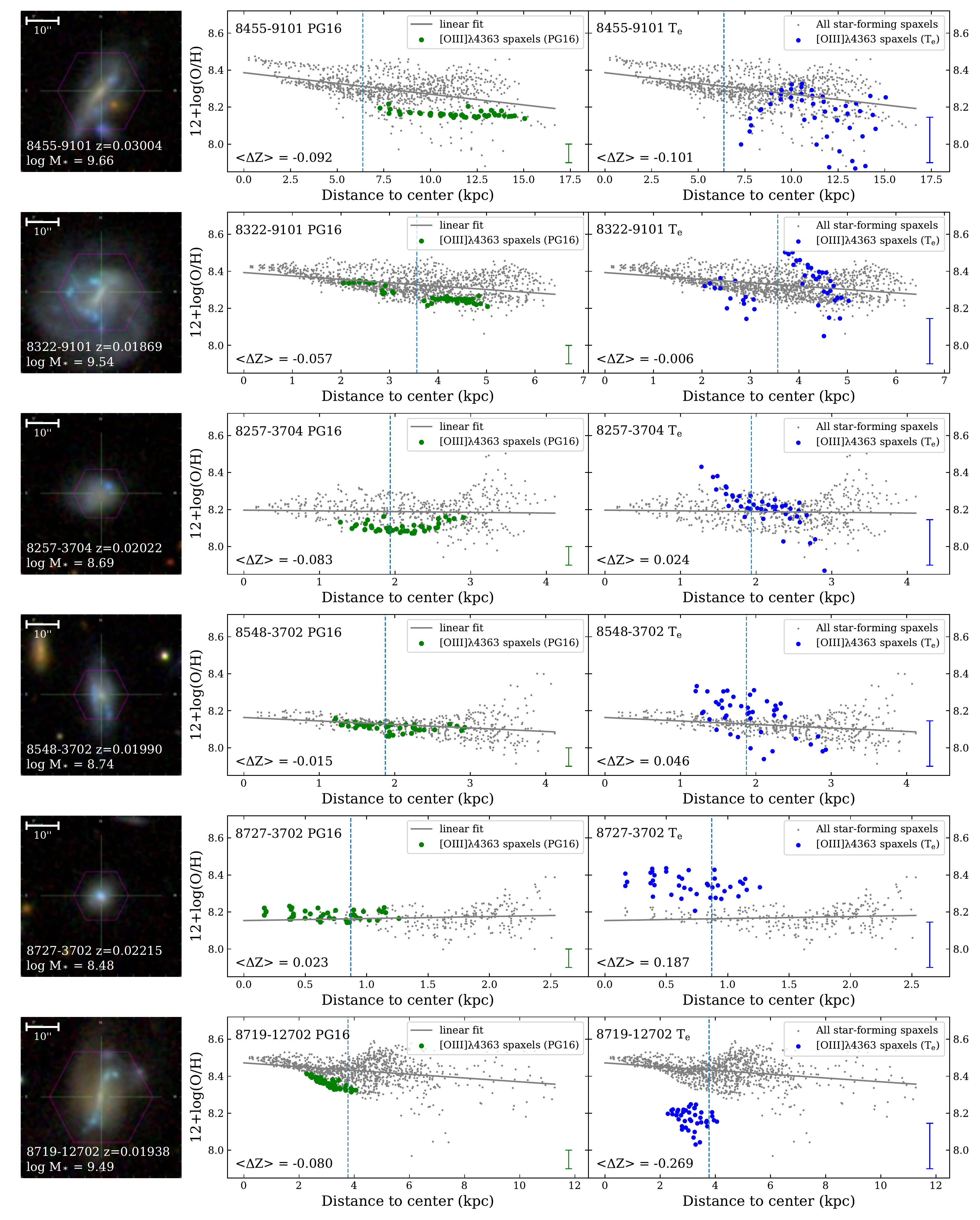}
	\caption{Same as Figure \ref{fig:metal_profile_0_5}, but the galaxies are ranked 11th to 16th in descending order of spaxels. \label{fig:metal_profile_11_17}}
\end{figure*}

\begin{figure*}[ht!]
	\centering
	\includegraphics[width=0.9\textwidth]{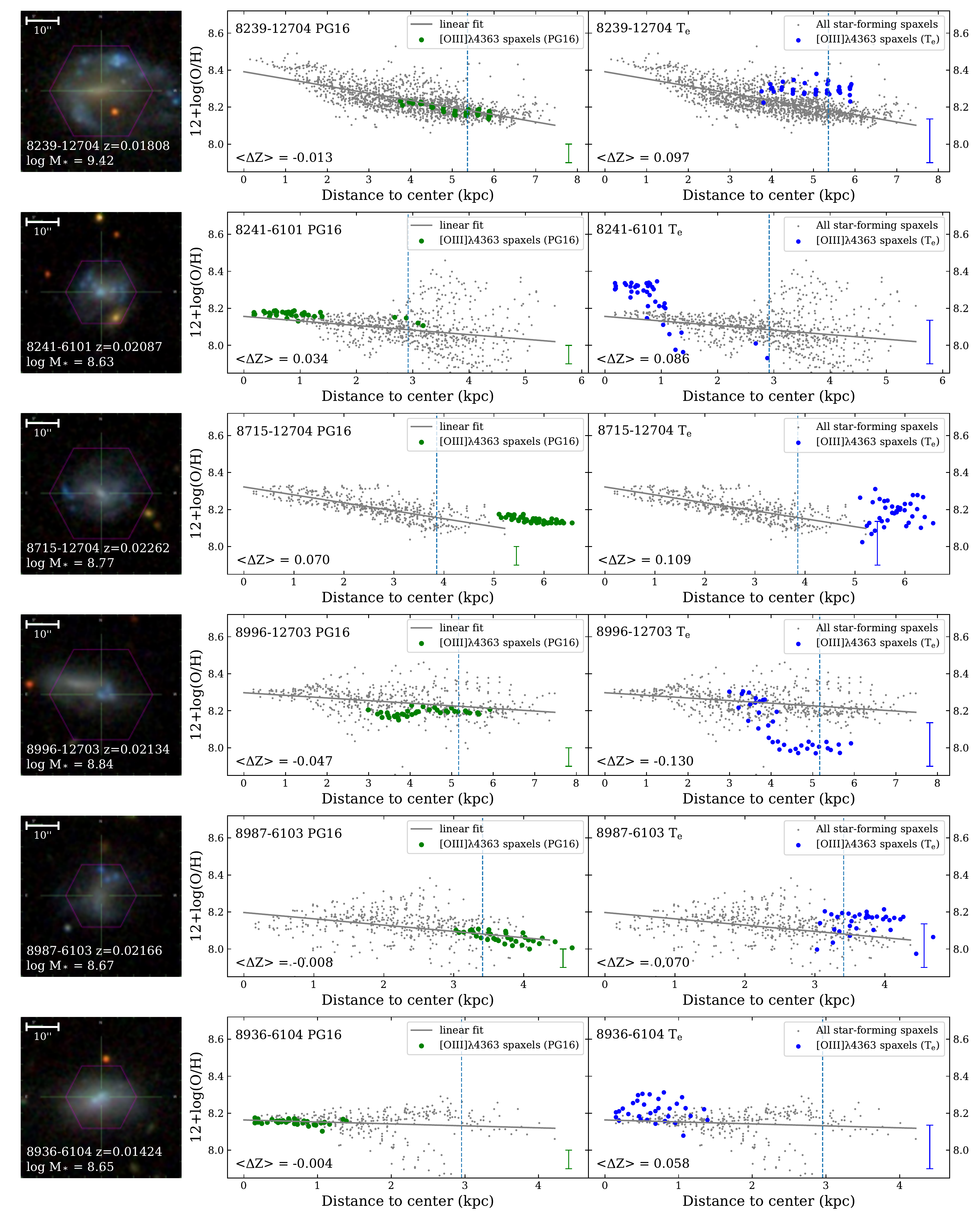}
	\caption{Same as Figure \ref{fig:metal_profile_0_5}, but the galaxies are ranked 17th to 22nd in descending order of spaxels. \label{fig:metal_profile_17_23}}
\end{figure*}



\end{document}